\documentclass[fleqn,usenatbib]{rasti}

\usepackage{mathptmx}
\usepackage[T1]{fontenc}

\DeclareRobustCommand{\VAN}[3]{#2}
\let\VANthebibliography\thebibliography
\def\thebibliography{\DeclareRobustCommand{\VAN}[3]{##3}\VANthebibliography}

\usepackage{amsmath}
\usepackage{amssymb}      
\usepackage{graphicx}
\usepackage{subcaption}   
\usepackage{booktabs}
\usepackage{array}
\usepackage{tabularx}
\usepackage{colortbl}     
\usepackage{longtable}
\usepackage{pdflscape}    
\usepackage{enumitem}
\usepackage{textcomp}
\usepackage{microtype}
\usepackage{hyperref}
\usepackage{orcidlink}    

\graphicspath{{figures/}}

\newcommand{\Y}{\checkmark}
\newcommand{\N}{\textendash}

\title[PICs for the Habitable Worlds Observatory]{Photonic Integrated
Circuits for the Habitable Worlds Observatory: Science Drivers, Material
Platforms, Arrayed-Waveguide Spectrographs, and a Space-Qualification
Roadmap}

\author[K. Madhav]{
  Kalaga Madhav\orcidlink{0000-0002-6711-7137}$^{1}$\thanks{
    Corresponding author. E-mail: \href{mailto:kmadhav@aip.de}{kmadhav@aip.de}}
  \\
  $^{1}$Astrophotonics (innoFSPEC), Leibniz Institute for Astrophysics
  Potsdam (AIP), An der Sternwarte 16, 14482 Potsdam, Germany
}

\date{Accepted XXX. Received YYY; in original form ZZZ}
\pubyear{\the\year{}}

\begin{document}
\label{firstpage}
\pagerange{\pageref{firstpage}--\pageref{lastpage}}
\maketitle

\begin{abstract}
NASA's Habitable Worlds Observatory (HWO) will require ultraviolet,
optical, and near-infrared instruments that are simultaneously compact,
mechanically and thermally stable, high-throughput, and replicable at
large channel counts. Space-qualified photonic integrated circuits (PICs)
and optical fibres are key platforms that can deliver these properties:
they manipulate light at the diffraction limit within micron-scale
single-mode waveguide circuits. An astrophotonic instrument fully guides
the starlight from focal plane to detector, eliminating scatter and ghosts
and offering excellent stability with no moving parts. This paper examines how
astrophotonics can address the principal science drivers of HWO through
four topics: (I) the HWO observing modes best implemented with photonic
instruments; (II) the waveguide and fibre materials that can span the
demanding 100\,nm--2.5\,\textmu{}m HWO wavelength range; (III) the
potential of arrayed-waveguide-grating (AWG) spectrographs to reach the
resolving powers and throughputs HWO science requires; and (IV) the steps
needed to space-qualify PICs and fibres against radiation, thermal
cycling, vacuum, and launch loads. For each topic, concrete technical
requirements are extracted and the current Technology Readiness Level
(TRL) is assessed. The main contribution of this paper is to unify these
science drivers, candidate-aperture photon budgets, UV-to-near-infrared
material platforms, AWG architectures, and space-qualification
requirements within a single photon-budget framework that pairs each
claimed benefit with a measurable requirement, a TRL, and an environmental
test that can retire it. A phased roadmap advances the critical photonic
components from their present TRL~2--5, depending on platform and
application, to the TRL~6 needed at HWO's instrument-definition gate.
\end{abstract}

\begin{keywords}
instrumentation: spectrographs -- instrumentation: photon collectors --
space vehicles: instruments -- techniques: spectroscopic --
methods: laboratory: solid state
\end{keywords}

\section{Introduction}
\label{sect:intro}

The 2020 Decadal Survey \emph{Pathways to Discovery in Astronomy and
Astrophysics for the 2020s}~\citep{Astro2020} recommended development of the
Habitable Worlds Observatory (HWO): an approximately six-metre
ultraviolet/optical/near-infrared (UVOIR) space telescope whose primary
purpose is the direct imaging and spectroscopic characterization of
potentially habitable exoplanets, and which must also deliver a broad
general-astrophysics capability. To develop that scope, NASA formed the
Science, Technology, Architecture Review Team (START), which invited the
community into working groups to explore the potential discovery space; the
process produced seventy science cases, documented as Science Case Development
Documents (SCDDs) and addressing twenty-seven of the thirty science questions
and discovery areas identified by
Astro2020~\citep{Dressing2026SCDD,HWO25Foreword}. Those documents are
collected in the HWO25 Proceedings~\citep{ASP542}.

A recurring conclusion across the SCDD library is that HWO must combine, on a
single platform, instrument capabilities conventionally distributed across
several bulk-optic facilities: ultraviolet access to wavelengths as short as
$\sim\!100\,\mathrm{nm}$; spectral resolving powers from $R\!\sim\!100$ to
$R\!\gtrsim\!10^{5}$; multi-object and integral-field formats with
$10^{2}$--$10^{4}$ apertures; coronagraphic contrasts approaching one part in
ten billion; and centimetre-per-second extreme-precision radial velocity
(EPRV). Each of these capabilities, implemented in conventional optics,
carries a penalty in volume, mass, mechanical complexity, thermal
sensitivity, or per-channel cost, penalties that are especially acute for
a space mission, where mass and stability budgets are unforgiving.

Photonic integrated circuits (PICs) offer a different implementation pathway.
A PIC routes and manipulates light within single-mode waveguides patterned
lithographically onto a chip of order one square centimetre. Because the light
is guided rather than propagated through free space, a PIC instrument has no
moving parts, occupies a millimetre-to-centimetre-scale volume, is intrinsically
stable against mechanical and thermal perturbation, and, critically for a
multi-channel instrument, can be replicated at wafer scale, enabling channel
counts of $10^{3}$--$10^{4}$ within a chip footprint of order cm$^{2}$, a
capability not previously demonstrated with ruled-grating spectrographs.
These properties are a strong match to the HWO instrumentation challenge. The application of integrated photonics
and specialty fibres to astronomy, the field of
astrophotonics~\citep{Bland2009Astrophotonics,Minardi2021Review}, has produced
on-sky demonstrations, most prominently the photonic beam combiner of the
ESO/GRAVITY instrument~\citep{GRAVITY2017}, and a broad community technology
assessment~\citep{Jovanovic2023Roadmap}.

This paper focuses on PICs and optical fibres as enabling technologies for HWO, organised around the four topics set out in the abstract: the observing modes best implemented with photonic instruments, the materials that span the HWO wavelength range, the AWG spectrographs that provide the required resolving power and throughput, and the steps to space-qualify these components. Section~\ref{sect:drivers} distils the HWO science drivers into instrument requirements and identifies the observing modes best served by photonic implementations, and Section~\ref{sect:photon-budget} quantifies the photon budget delivered by each of the three candidate telescope apertures, establishing why throughput is decisive. Section~\ref{sect:pic-applications} develops topic-I, mapping each observing mode onto a concrete PIC application. Section~\ref{sect:materials} takes up topic-II, the waveguide and fibre material platforms that together span the 100\,nm--2.5\,\textmu m HWO wavelength range. Section~\ref{sect:awg} takes up topic-III, the potential of arrayed-waveguide-grating (AWG) spectrographs to reach the resolving powers and throughputs HWO science requires. Section~\ref{sect:space} takes up topic-IV, the space qualification of PICs and fibres against radiation, thermal cycling, vacuum, and launch loads. Section~\ref{sect:trl} then consolidates, across all four topics, the key technical requirements and Technology Readiness Levels (TRLs); Section~\ref{sect:roadmap} presents a phased development roadmap; and Section~\ref{sect:conclusions} concludes.

\section{HWO Science Drivers and Instrument Requirements}
\label{sect:drivers}

\subsection{Distillation of the Science Case Library}

The science cases were developed by four community working groups, Living
Worlds, Growth of Galaxies, Solar Systems in Context, and Evolution of the
Elements, whose top-level instrument requirements are summarised in
Table~\ref{tab:drivers}. Aggregated across the seventy science cases, access
to ultraviolet wavelengths is the single most widespread requirement: eighty-three
percent of the cases need data shortward of 400\,nm and twenty-six percent
extend below 100\,nm, while twenty-six percent require observations at
wavelengths of 2000\,nm or longer. Spectroscopy is required by eighty-seven
percent of the cases and photometry by thirty percent, with high-contrast and
polarimetric capabilities needed by thirty-four and twenty-seven percent
respectively~\citep{Dressing2026SCDD}. The observing modes therefore divide
between imaging, long-slit, integral-field, multi-object, and high-contrast
spectroscopy.

\begin{table*}
\caption{HWO science drivers and their top-level instrument requirements,
grouped by working group. Modes: Coro = coronagraphy; IFS = integral-field
spectroscopy; MOS = multi-object spectroscopy; EPRV = extreme-precision
radial velocity.}
\label{tab:drivers}
\begin{center}
\footnotesize
\begin{tabularx}{\textwidth}{|l|l|l|l|X|}
\hline
\rule[-1ex]{0pt}{3.5ex}
\textbf{Working group} & \textbf{Wavelengths} & \textbf{Resolution $R$} &
\textbf{Mode} & \textbf{Representative science driver} \\
\hline\hline
\rule[-1ex]{0pt}{3.5ex}
Living Worlds &
$0.2$--$1.8\,$\textmu m &
\,70--500 &
Coro+IFS &
Earth-analog biosignatures (O$_2$, O$_3$, H$_2$O)~\protect\citep{Parenteau2025LW} \\
\hline
\rule[-1ex]{0pt}{3.5ex}
Growth of Galaxies &
$94$--$350\,$nm &
$\,10^{3}$--$10^{5}$ &
MOS+IFS &
circumgalactic-medium absorption mapping~\protect\citep{Borthakur2025DiskCGM,Burchett2025CGM} \\
\hline
\rule[-1ex]{0pt}{3.5ex}
Solar Systems in Context &
$0.1$--$2.5\,$\textmu m &
$\,70$--$3\!\times\!10^{5}$ &
Coro+IFS+EPRV &
exoplanet atmospheres; terrestrial-planet masses~\protect\citep{Lopez2025HighRes,Brandt2025Masses} \\
\hline
\rule[-1ex]{0pt}{3.5ex}
Evolution of the Elements &
$94\,$nm--$1.8\,$\textmu m &
$\,10^{3}$--$10^{5}$ &
IFS+MOS &
stellar magnetic fields; chemical enrichment~\protect\citep{Strugarek2025MagFields} \\
\hline
\end{tabularx}
\end{center}
\end{table*}

\subsection{Six Observing-Mode Drivers Suited to Photonic Implementation}

From the SCDD requirements, six observing-mode capabilities were identified for
which a photonic implementation is either advantageous or, in practice,
enabling:

\begin{enumerate}[leftmargin=2em,itemsep=2pt]
\item \textbf{Diffraction-limited spectroscopy behind the coronagraph.}
Direct imaging of Earth analogs requires a contrast approaching one part in
ten billion at inner working angles of order $3\,\lambda/D$, followed by
$R\!\sim\!100$--$500$ spectroscopy of the planet light~\citep{Parenteau2025LW}.
Because the coronagraph delivers a near-diffraction-limited beam, the planet
light couples efficiently into single-mode waveguides, the native input
of a PIC.

\item \textbf{High-resolution cross-correlation spectroscopy (HRCCS).}
Retrieval of molecular biosignatures benefits from $R\!\sim\!10^{5}$ over a
broad UV-to-near-infrared band~\citep{Lopez2025HighRes}. This is the most
demanding spectrograph requirement in the SCDD library.

\item \textbf{Multi-object UV spectroscopy.}
Circumgalactic-medium and reionization science requires far-ultraviolet
multi-object spectroscopy with $10^{3}$--$10^{4}$ apertures at
$R\!\sim\!10^{4}$--$10^{5}$~\citep{Borthakur2025DiskCGM,Citro2025LyC,RomanDuval2025Dust}.
Replicable PIC channels are well matched to this multiplexing requirement.

\item \textbf{Integral-field spectroscopy at UV and optical wavelengths.}
Spatially resolved spectroscopy of the circumgalactic medium, AGN outflows,
and ocean-world atmospheres requires UV/optical integral-field units at
$R\!\sim\!3000$--$30{,}000$~\citep{Burchett2025CGM,LustigYaeger2025Ocean}.

\item \textbf{Extreme-precision radial velocity.}
Determining the masses of HWO's directly-imaged habitable-zone targets
requires RV precision at the centimetre-per-second level over decadal
baselines~\citep{Brandt2025Masses,Stark2014Yield}, which in turn requires an
ultra-stable spectrograph and a stable wavelength calibrator.

\item \textbf{High-resolution UV spectropolarimetry.}
Measurement of stellar and exoplanetary magnetic fields requires
$R\!\sim\!10^{5}$ UV spectropolarimetry at the $10^{-5}$ polarization
level~\citep{Strugarek2025MagFields}.
\end{enumerate}

These six drivers set the photon-budget analysis of
Section~\ref{sect:photon-budget}, motivate the PIC applications of
Section~\ref{sect:pic-applications}, set the material requirements of
Section~\ref{sect:materials}, and define the spectrograph specifications of
Section~\ref{sect:awg}.

\section{Photon Budgets for the Three Exploratory Analytic Cases}
\label{sect:photon-budget}

The HWO telescope aperture is not yet fixed: NASA's architecture studies
currently use three Exploratory Analytic Cases (EACs)~\citep{Liu2026EAC}. EAC~1 is
a 6.0\,m inscribed / 7.2\,m circumscribed off-axis segmented hexagon; EAC~2 is
a 6.0\,m off-axis unobscured aperture; and EAC~3 is an 8.0\,m on-axis
segmented aperture. Because every astrophotonic component discussed in this
paper acts on a finite, and, for the direct-imaging drivers, severely
photon-starved signal, the photon budget delivered by each EAC sets the
quantitative context for the throughput requirements of
Sections~\ref{sect:materials}--\ref{sect:awg}. For
each science driver and each EAC, the photons delivered per spectral
resolution element is quantified.

\subsection{Effective Collecting Areas}

Table~\ref{tab:eac} lists the effective collecting areas adopted here. EAC~1
is treated as a regular hexagon of 7.2\,m circumscribed diameter with a
three-percent allowance for inter-segment gaps; EAC~2 as an unobscured
6.0\,m circle; and EAC~3 as an 8.0\,m circle with a thirteen-percent central
obscuration representative of an on-axis segmented design. These values are
accurate to roughly $\pm15\,\%$ and are intended to expose the relative
scaling between cases rather than to substitute for a mission throughput
model.

\begin{table}
\caption{Effective collecting areas adopted for the three HWO Exploratory
Analytic Cases (EACs). Areas are approximate ($\pm15\,\%$) and depend on the
final segmentation and obscuration.}
\label{tab:eac}
\begin{center}
\footnotesize
\begin{tabularx}{\linewidth}{|l|X|c|c|}
\hline
\rule[-1ex]{0pt}{3.5ex}
\textbf{Case} & \textbf{Aperture description} & \textbf{Obscuration} &
\textbf{$A_{\mathrm{eff}}$ (m$^2$)} \\
\hline\hline
\rule[-1ex]{0pt}{3.5ex}
EAC 1 & 6.0\,m inscribed / 7.2\,m circumscribed, off-axis segmented hexagon &
none & $\approx 33$ \\
\hline
\rule[-1ex]{0pt}{3.5ex}
EAC 2 & 6.0\,m circular, off-axis & none & $\approx 28$ \\
\hline
\rule[-1ex]{0pt}{3.5ex}
EAC 3 & 8.0\,m circumscribed, on-axis segmented & $\sim$13\,\% & $\approx 44$ \\
\hline
\end{tabularx}
\end{center}
\end{table}

\subsection{Photon-Budget Method}

For a source of AB magnitude $m_{\mathrm{AB}}$, the photon flux density is
$\Phi_\lambda = (5.48\times10^{6}/\lambda_{\mathrm{\AA}})\,
10^{-0.4\,m_{\mathrm{AB}}}$ photons\,s$^{-1}$\,cm$^{-2}$\,\AA$^{-1}$. The
photon rate collected per spectral resolution element is
\begin{equation}
\label{eq:photonrate}
\dot{N} = \Phi_\lambda \; A_{\mathrm{eff}} \; \eta \;
          \frac{\lambda}{R} \,
\end{equation}
where $A_{\mathrm{eff}}$ is the effective collecting area, $\eta$ the
end-to-end throughput (telescope $\times$ instrument $\times$ detector, and
$\times$ coronagraph core throughput where applicable), and $\lambda/R$ the
width of one resolution element. The photon-noise-limited integration time
to a signal-to-noise ratio $\mathrm{SNR}$ per resolution element is
$t = \mathrm{SNR}^{2}/\dot{N}$; this is a lower bound, since zodiacal and
exozodiacal background and detector noise lengthen realistic exposures by
factors of two to ten.

A useful consequence of Eq.~(\ref{eq:photonrate}) follows from the AB system
being defined per unit frequency: at fixed $m_{\mathrm{AB}}$ and $R$, the
product $\Phi_\lambda\,(\lambda/R)$ is independent of wavelength. The
wavelength structure of the photon budget is therefore carried almost
entirely by the throughput $\eta(\lambda)$ and by the source spectrum
$m_{\mathrm{AB}}(\lambda)$, and $\eta(\lambda)$ is precisely where
astrophotonic material losses, fibre transmission, and coupling efficiency
enter. The photon budget is, in this sense, a direct measure of the cost of
astrophotonic throughput.

\subsection{Photon Budgets by Science Driver}

Table~\ref{tab:photonbudget} applies Eq.~(\ref{eq:photonrate}) to the six
science drivers of Section~\ref{sect:drivers}, using representative target
brightnesses, resolving powers, and throughputs. The throughput values are
deliberately conservative end-to-end estimates: $\eta\!\approx\!0.03$--$0.05$
in the far- and near-ultraviolet, where coating, detector, and PIC-material
losses compound; $\eta\!\approx\!0.07$--$0.11$ for coronagraph-fed and NUV
modes; and $\eta\!\approx\!0.20$ for direct (non-coronagraphic) optical
spectroscopy.

\begin{table*}
\caption{Photon budget per spectral resolution element for the six HWO
science drivers and the three EACs. $\dot{N}$ is in photons per hour;
$t$ is the photon-noise-limited time to the benchmark SNR (lower bound).
Exo-Earth targets are at 10\,pc in reflected light. HRCCS rows are per single
resolution element; cross-correlation over $10^{3}$--$10^{4}$ lines shortens
the effective detection time by that factor. The throughput $\eta$ is a single
representative value at the listed wavelength; actual $\eta(\lambda)$ varies
substantially across the full HWO band (UV $\lesssim0.1$, visible $\sim0.07$--$0.20$,
NIR $\sim0.15$--$0.25$) owing to wavelength-dependent coating losses, waveguide
material absorption, and detector quantum efficiency (see Table~\ref{tab:throughput}).
Rows spanning a wavelength range use the $\eta$ value representative of the
dominant wavelength sub-band listed.}
\label{tab:photonbudget}
\begin{center}
\scriptsize
\begin{tabularx}{\textwidth}{|X|c|c|c|c|c|c|c|c|}
\hline
\rule[-1ex]{0pt}{3.2ex}
\textbf{Science driver} & \textbf{$m_{\mathrm{AB}}$} &
\textbf{$\lambda$} & \textbf{$R$} & \textbf{$\eta$} &
\textbf{EAC\,1} & \textbf{EAC\,2} & \textbf{EAC\,3} &
\textbf{$t$ (E1/E2/E3)} \\
 & & (nm) & & & (ph/hr) & (ph/hr) & (ph/hr) & SNR set \\
\hline\hline
\rule[-1ex]{0pt}{3ex}
Exo-Earth biosignatures, UV (O$_3$) & 29.5 & 250 & 140 & 0.03 &
2.2 & 1.9 & 2.9 & 1.9/2.2/1.4\,d \\
\hline
\rule[-1ex]{0pt}{3ex}
Exo-Earth biosignatures, VIS--NIR (O$_2$, H$_2$O) & 29.5 & 550--1000 & 140 & 0.07 &
5.1 & 4.4 & 6.8 & 20/23/15\,h \\
\hline
\rule[-1ex]{0pt}{3ex}
HRCCS atmospheres, per res.\ elt & 29.5 & 550--1500 & $10^5$ & 0.06 &
$6\!\times\!10^{-3}$ & $5\!\times\!10^{-3}$ & $8\!\times\!10^{-3}$ &
680/785/510\,d \\
\hline
\rule[-1ex]{0pt}{3ex}
Multi-object UV spectroscopy, bright & 21.0 & 150 & $10^5$ & 0.05 &
12.8 & 11.1 & 17.2 & 7.8/9.0/5.8\,h \\
\hline
\rule[-1ex]{0pt}{3ex}
Multi-object UV spectroscopy, faint & 24.5 & 150 & $2\!\times\!10^4$ & 0.05 &
2.6 & 2.2 & 3.4 & 1.6/1.9/1.2\,d \\
\hline
\rule[-1ex]{0pt}{3ex}
UV/optical integral-field spectroscopy & 23.0 & 300 & 5000 & 0.11 &
89 & 77 & 120 & 1.1/1.3\,h, 50\,min \\
\hline
\rule[-1ex]{0pt}{3ex}
Extreme-precision radial velocity & 5.0 & 550 & $1.5\!\times\!10^5$ & 0.20 &
$8.6\!\times\!10^7$ & $7.4\!\times\!10^7$ & $1.2\!\times\!10^8$ &
0.4/0.5/0.3\,s \\
\hline
\rule[-1ex]{0pt}{3ex}
UV spectropolarimetry & 8.0 & 200 & $10^5$ & 0.05 &
$2.0\!\times\!10^6$ & $1.8\!\times\!10^6$ & $2.7\!\times\!10^6$ &
2.7/3.1/2.0\,min \\
\hline
\end{tabularx}
\end{center}
\end{table*}

The benchmark SNR is 10 per resolution element for the spectroscopic drivers,
100 for radial velocity (a proxy for the cm\,s$^{-1}$ photon-noise floor),
and 300 for spectropolarimetry (a proxy for a $10^{-3}$ polarization
measurement). Three regimes are evident. The direct-imaging drivers (exo-Earth
biosignatures and HRCCS) are profoundly photon-starved, a few photons per
hour per resolution element, and for HRCCS, a few thousandths of a photon
per hour, so that biosignature spectra require tens of hours per band and
HRCCS is feasible only because cross-correlation coadds thousands of lines.
The ultraviolet survey drivers (multi-object spectroscopy) occupy an
intermediate regime of hours to a few days. The bright-target drivers
(radial velocity, spectropolarimetry) are photon-rich; there the budget
question is not the number of photons but whether the instrument is stable
enough to reach the photon-noise floor; the wavelength stability and
polarization control that motivate the PIC implementations of
Section~\ref{sect:pic-applications}.

\subsection{Implications for Astrophotonic Throughput}
\label{subsect:throughput-implications}

Two quantitative conclusions connect the photon budget to the rest of this
paper. First, the EAC scaling is modest but real: relative to EAC~2, EAC~1
collects $1.16\times$ and EAC~3 collects $1.55\times$ the photons, so
photon-noise-limited exposure times scale by $0.87\times$ and $0.65\times$
respectively. For the photon-starved exo-Earth drivers this is the difference
between roughly 15 and 23 hours per spectral band per target which,
multiplied across the target list, directly sets how many potentially
habitable worlds HWO can characterise within its lifetime.

Second, and central to this paper: the end-to-end throughput $\eta$ enters
Eq.~(\ref{eq:photonrate}) on the same footing as the collecting area
$A_{\mathrm{eff}}$. The astrophotonic spectrograph throughput budget of
Section~\ref{sect:awg} ranges from about 27\,\% unoptimised to about 79\,\%
optimised (optical-only efficiency, excluding detector QE), a factor of three.
In the photon budget, that factor of three is indistinguishable from a factor
of three in collecting area: an unoptimised photonic instrument on EAC~3
delivers barely half the photons of an optimised one on EAC~2, even though
EAC~3 has the larger aperture. Conversely, raising astrophotonic throughput
from the unoptimised to the optimised value recovers as much sensitivity as
moving from a 6\,m to a 10\,m aperture, at a small fraction of the mass. This
equivalence is
the quantitative core of the case for astrophotonics: because HWO is
photon-starved in exactly the science that defines the mission, photonic
throughput is not an incremental optimisation but a lever comparable to the
choice of aperture itself.

\section{Photonic Integrated Circuits for HWO Science Drivers}
\label{sect:pic-applications}

This section maps each of the six observing-mode drivers onto a concrete PIC
implementation. The common architectural thread is that a PIC instrument
accepts single-mode (or few-mode) inputs, performs the required optical
function, dispersion, beam combination, filtering, or calibration
within waveguides, and delivers light to an integrated or butt-coupled
detector.

\subsection{Mapping astrophotonic functions to HWO telescope-level technical risks}
\label{subsec:tech-telescope-risk-map}

The value proposition of photonic integrated circuits (PICs) for the Habitable Worlds Observatory (HWO) is not only miniaturisation of spectrographs, but also risk reallocation at the telescope--instrument interface.  A bulk spectrograph treats many observatory perturbations such as, pointing jitter, thermal breathing, structural drift, mechanism repeatability, contamination, and detector-area exposure, as optical-layout or calibration problems.  Guided-wave architectures can instead move part of this burden into a compact, replicated, thermally co-located, and mechanically static photonic layer.  
Table~\ref{tab:techissues} maps
the five astrophotonic technology families (PICs and AWG spectrographs,
optical fibres, photonic integral-field units, microlens arrays, and
photonic lanterns) against ten space-telescope issue classes, distinguishing,
for each pairing, the \emph{potential} (how the technology mitigates the
issue) from the \emph{challenge} (how it remains vulnerable to, or itself
introduces, the issue), and summarising the net effect on the photon
budget. The recurring pattern is that guided-wave technologies trade a
modest, calibratable, and predictable loss term for the removal of a larger,
variable, or harder-to-calibrate one - a favourable exchange for the
photon-starved science of Section~\ref{sect:photon-budget}.

\clearpage
\onecolumn
\begin{landscape}
\begingroup
\setlength{\textwidth}{9in}\setlength{\linewidth}{9in}
\setlength{\LTcapwidth}{9in}
\centering
\scriptsize
\renewcommand{\arraystretch}{1.02}
\begin{longtable}{|p{2.5cm}|p{6.7cm}|p{6.7cm}|p{4.2cm}|}
\caption{Astrophotonic technologies versus space-telescope technical issues.
Issue classes are drawn from a system-level taxonomy of space-telescope
failure modes and degradation mechanisms. For each technology--issue pairing,
the \emph{potential} is how the technology mitigates the issue and the
\emph{challenge} is how it remains vulnerable to, or introduces, the issue.}
\label{tab:techissues}\\
\hline
\rowcolor[gray]{0.80}
\textbf{Technology} & \textbf{Potential (how it mitigates / helps)} &
\textbf{Challenge (how it is vulnerable / what it introduces)} &
\textbf{Net effect on photon budget}\\
\hline
\endfirsthead
\multicolumn{4}{l}{\footnotesize\itshape Table~\thetable\ (continued)}\\[2pt]
\hline
\rowcolor[gray]{0.80}
\textbf{Technology} & \textbf{Potential (how it mitigates / helps)} &
\textbf{Challenge (how it is vulnerable / what it introduces)} &
\textbf{Net effect on photon budget}\\
\hline
\endhead
\hline
\multicolumn{4}{r}{\footnotesize\itshape continued on next page}\\
\endfoot
\hline
\endlastfoot

\rowcolor[gray]{0.90}\multicolumn{4}{|l|}{\textbf{Fine-guidance jitter /
slit-coupling loss} \ (Pointing \& attitude control)}\\
\hline
Photonic lanterns &
A few-mode lantern accepts a jittering, partially-corrected beam without the
coupling collapse a single-mode fibre suffers; it can also act as a
focal-plane tip-tilt / wavefront sensor. &
Throughput still falls if jitter drives the input beyond the lantern mode
count; UV lanterns are undemonstrated. &
Positive: stabilises the injection term that jitter would otherwise
modulate.\\
\hline
Optical fibres &
A fibre feed with mechanical scrambling decouples the spectrograph from
line-of-sight motion; azimuthal and radial scrambling stabilises the near-
and far-field. &
Focal-ratio degradation and residual modal noise; coupling still varies with
the PSF delivered to the fibre face. &
Positive: converts jitter-induced coupling loss into a smaller, calibratable
modal-noise term.\\
\hline
Microlens arrays &
A lenslet over the fibre core widens the effective acceptance and flattens
the coupling-versus-offset curve, reducing jitter sensitivity. &
Adds reflective and anti-reflection-coating losses; lenslet-to-fibre
alignment tolerance is tight. &
Positive: raises and flattens the injection efficiency.\\
\hline

\rowcolor[gray]{0.90}\multicolumn{4}{|l|}{\textbf{Spectral-format drift;
wavelength-calibration drift} \ (Spectrographs; Calibration)}\\
\hline
PICs / AWG spectrographs &
No moving parts and a millimetre-scale, thermally co-located circuit make
the dispersion solution intrinsically more stable than a bulk echelle on a
large bench. &
Residual thermo-optic drift of the waveguide index still shifts the
solution; active temperature control or on-chip athermalisation is needed. &
Positive: a smaller, more predictable drift term; less calibration overhead
lost to dead time.\\
\hline
Optical fibres &
A fibre feed isolates the disperser from the telescope thermal breathing and
structural relaxation that move bulk spectral formats. &
Fibre stress and temperature changes alter modal content and hence the
line-spread function. &
Positive: removes a major drift source; introduces a smaller modal-noise
term.\\
\hline

\rowcolor[gray]{0.90}\multicolumn{4}{|l|}{\textbf{Grating-wheel /
disperser-mechanism failure; filter-wheel failure} \ (Mechanisms / moving
parts)}\\
\hline
PICs / AWG spectrographs &
An on-chip AWG has no moving disperser, so this single-point
mechanism-failure mode is removed; spectral selection can be done by routing
rather than by a wheel. &
Wavelength agility now requires either multiple chips or tunable
(thermo-optic) elements, which add electrical complexity. &
Neutral-to-positive: trades a mechanism risk for static, replicated
channels.\\
\hline

\rowcolor[gray]{0.90}\multicolumn{4}{|l|}{\textbf{Mass, volume, and stiffness
constraints} \ (Mission design / system engineering)}\\
\hline
PICs / AWG spectrographs &
Centimetre-scale chips replace metre-scale spectrograph benches; wafer
replication adds channels at marginal mass, decisive for multi-object
instruments. &
Hybrid packaging, the electrical harness, and detector integration reclaim
some of the mass and volume saving. &
Positive: relaxes the mass budget, allowing more collecting area or shielding
within a fixed launch mass.\\
\hline
Photonic IFUs &
A guided-optics reformatter is far more compact than a bulk image slicer for
the same spaxel count. &
Throughput cost of many interfaces; UV operation is immature. &
Positive on mass; must be watched on throughput.\\
\hline

\rowcolor[gray]{0.90}\multicolumn{4}{|l|}{\textbf{Total ionising dose;
displacement damage; solar energetic particles} \ (Radiation / space
environment)}\\
\hline
Optical fibres &
Hollow-core anti-resonant fibres guide light in vacuum, largely sidestepping
the radiation-induced absorption and solarisation that darken solid silica. &
Solid UV-grade silica and fluoride fibres darken under dose, worst in the UV;
each fibre type must be dose-tested. &
Mixed: hollow-core protects the budget; solid fibres erode it over mission
life.\\
\hline
PICs / AWG spectrographs &
A small interaction volume and crystalline wide-bandgap platforms (AlN,
Al$_2$O$_3$) can be comparatively radiation-tolerant. &
Radiation-induced absorption and colour-centre formation are
material-specific and largely uncharacterised for astrophotonic platforms. &
Unknown until tested: a key qualification gap.\\
\hline

\rowcolor[gray]{0.90}\multicolumn{4}{|l|}{\textbf{Molecular contamination /
outgassing; water-ice deposition} \ (Contamination)}\\
\hline
PICs / AWG spectrographs &
A guided, enclosed light path exposes little cold optical surface for
contaminants to settle on; outgassing index-matching gels can be eliminated
by solid bonds. &
Fibre-to-chip bonds must be achieved without outgassing adhesives; chip
facets remain contamination-sensitive. &
Positive: a smaller exposed surface protects throughput over mission life.\\
\hline
Optical fibres &
An enclosed fibre core is not exposed to deposition along its length. &
Fibre end faces and connectors remain contamination-sensitive. &
Positive for the transport path; end faces still need cleanliness control.\\
\hline

\rowcolor[gray]{0.90}\multicolumn{4}{|l|}{\textbf{Coronagraph wavefront
instability; speckle drift; polarization leakage} \ (High-contrast exoplanet
instruments)}\\
\hline
Photonic lanterns &
A single-mode output spatially filters residual speckle; a lantern wavefront
sensor measures low-order aberrations at the science focal plane with zero
non-common-path error. &
Speckle removed by spatial filtering is rejected, not recovered, a real
(if calibratable) throughput loss on the planet. &
Positive for contrast stability; the planet photon is taxed by the
single-mode coupling efficiency.\\
\hline
PICs / AWG spectrographs &
A polarisation-split AWG can analyse the two states separately, turning
polarization leakage into a measured quantity. &
Waveguide birefringence itself must be controlled or it degrades resolution. &
Neutral: enables polarization-aware operation at a modest throughput cost.\\
\hline

\rowcolor[gray]{0.90}\multicolumn{4}{|l|}{\textbf{Detector cosmic rays;
persistence; bad pixels; read noise} \ (Detectors / focal plane)}\\
\hline
PICs / AWG spectrographs &
An on-chip disperser concentrates a spectrum onto a small, well-characterised
detector region and enables butt-coupled photon-counting detectors (e.g.\
MKIDs), reducing read noise and the cosmic-ray-affected area. &
Detector integration onto the chip is itself an immature,
CTE-mismatch-prone step. &
Positive: a smaller detector footprint sees fewer cosmic-ray hits and less
dark current.\\
\hline

\rowcolor[gray]{0.90}\multicolumn{4}{|l|}{\textbf{Scattered light from surface
roughness; stray light from off-axis sources} \ (Optics / telescope
wavefront)}\\
\hline
Optical fibres &
A single-mode fibre feed acts as a spatial filter that rejects scattered and
stray light not in the science mode. &
Light rejected is light lost; the rejection helps contrast, not raw
throughput. &
Positive for background and contrast; neutral-to-negative on raw photon
count.\\
\hline
Photonic lanterns &
Modal filtering rejects out-of-mode stray light before it reaches the
disperser. &
The same trade: rejected stray light is removed, not recovered. &
Positive for background-limited science.\\
\hline

\rowcolor[gray]{0.90}\multicolumn{4}{|l|}{\textbf{Crowding and source
confusion; undersampling} \ (Science operations)}\\
\hline
Photonic IFUs &
Contiguous spaxel sampling resolves blended sources and removes slit-loss
ambiguity in crowded fields. &
Finite spaxel count and reformatter throughput; UV operation is
undemonstrated. &
Positive: recovers photons otherwise lost to blending and slit losses.\\
\hline
Microlens arrays &
Lenslet sampling at or below the diffraction limit mitigates undersampling
and fixes the spatial sampling independently of the detector pixel grid. &
Chromatic focal-length variation across the HWO band; alignment tolerance. &
Positive: stabilises the spatial-sampling term of the budget.\\
\hline
\end{longtable}
\endgroup
\end{landscape}
\clearpage
\twocolumn

This mapping motivates a qualification strategy in which each photonic subsystem is evaluated not only as an isolated component but as a term in the observatory photon budget.  For HWO, the relevant figure of merit is therefore not simply chip insertion loss.  It is the time-dependent end-to-end throughput, line-spread-function stability, polarisation stability, background rejection, and calibration overhead after exposure to launch, radiation, vacuum, thermal cycling, contamination, and long-duration operations.  In this sense, astrophotonics should be treated as a system-level stabilising layer: it can reduce several first-order telescope and mechanism sensitivities, but only by introducing new material, packaging, and interface terms that must be explicitly qualified.

\begin{table}
\centering
\caption{Risk-to-test translation for photonic subsystems proposed for HWO.  This table can be used as a bridge between the science-driver discussion and the later space-qualification roadmap.}
\label{tab:hwo_risk_to_test_translation}
\footnotesize
\setlength{\tabcolsep}{4pt}
\renewcommand{\arraystretch}{1.15}
\begin{tabularx}{\columnwidth}{>{\raggedright\arraybackslash}p{0.24\columnwidth} X X}
\toprule
\textbf{Risk class} & \textbf{Photonic observables to track} & \textbf{Qualification test implication} \\
\midrule
Injection stability & Coupling efficiency, modal distribution, focal-ratio degradation, near-/far-field stability & Jittered-beam injection tests with representative HWO PSFs, thermal perturbations, and vibration-before/after comparison. \\
\addlinespace
Spectral stability & AWG wavelength solution, resolving power, line-spread function, polarisation splitting, thermo-optic coefficient & Vacuum thermal cycling and long-duration drift tests with comb or line-source calibration. \\
\addlinespace
Radiation endurance & Transmission loss, colour-centre formation, annealing behaviour, birefringence change, darkening in fibres and waveguides & Total-ionising-dose and displacement-damage campaigns on each material stack, with in-situ or pre/post optical metrology. \\
\addlinespace
Contamination and bonding & Facet transmission, scattering, adhesive outgassing, bond survival, connector cleanliness & ASTM/ECSS-style outgassing screening, bake-out, contamination exposure, and post-exposure insertion-loss and scatter measurements. \\
\addlinespace
Detector/interface integration & Coupling efficiency to detector, thermal-stress survival, pixel mapping, cosmic-ray-sensitive area, read-noise impact & Cryogenic thermal cycling, mechanical survivability, detector co-registration, and radiation/cosmic-ray susceptibility assessment. \\
\bottomrule
\end{tabularx}
\end{table}

The principal conclusion is that photonic integration changes the allocation of risk rather than removing it outright.  AWG spectrographs and photonic IFUs directly address mass, volume, mechanism count, spectral-format stability, and detector-area exposure.  Photonic lanterns, fibres, and microlens arrays address coupling stability, modal control, and spatial sampling.  However, the enabling technologies introduce qualification gaps in UV operation, radiation response, thermo-optic control, fibre--chip attachment, contamination control, polarisation stability, and detector hybridisation.  These gaps define the experimental content of a credible HWO photonic-qualification roadmap: every claimed system-level benefit should be paired with a measurable photon-budget term and an environmental test that can retire it.

\subsection{Diffraction-Limited Spectroscopy Behind the Coronagraph}

\begin{figure}
\centering
\includegraphics[width=0.95\linewidth]{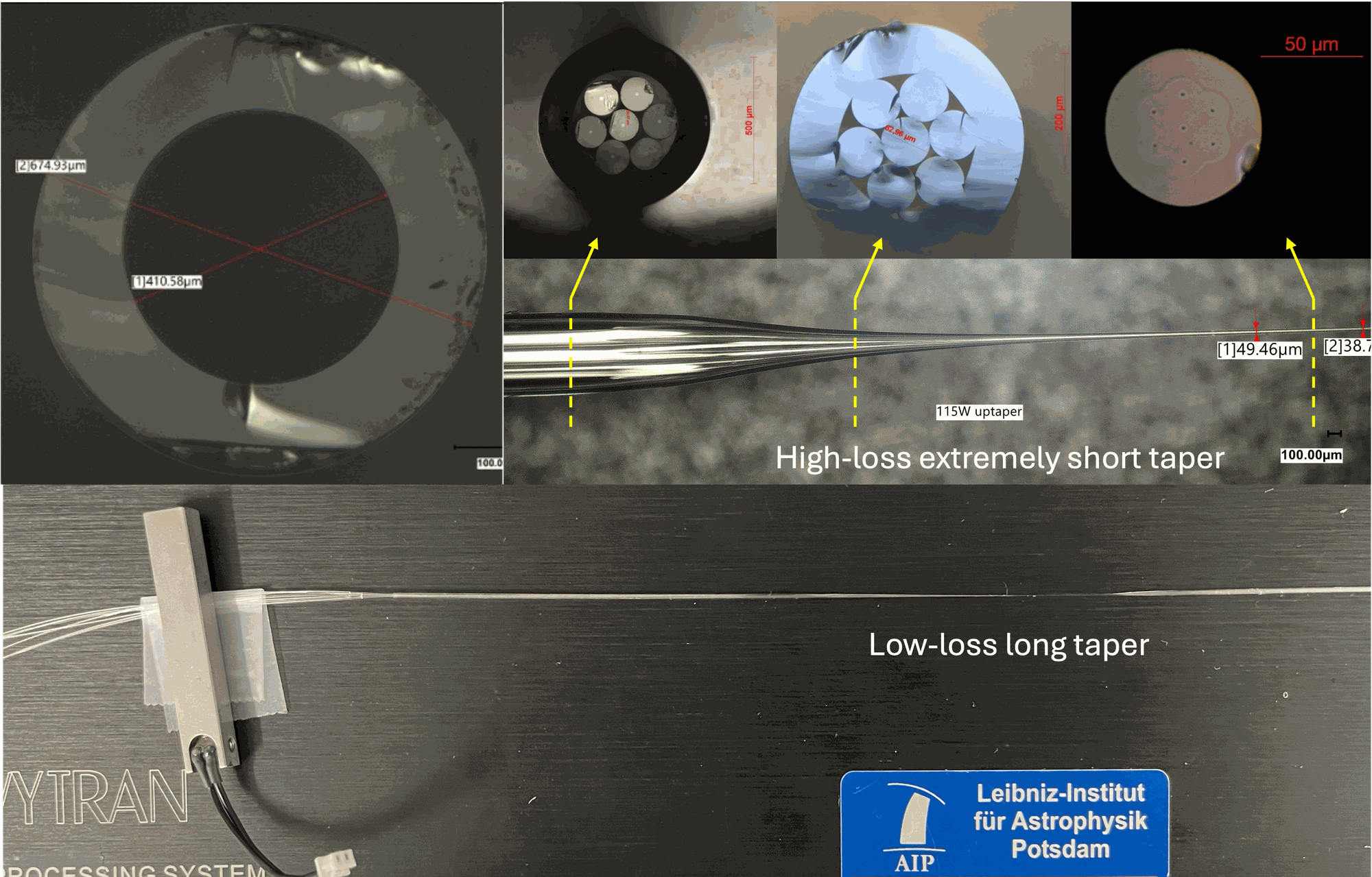}
\caption{Photonic lantern, the critical coupling interface between the coronagraphic focal plane and the single-mode AWG spectrograph. The panels show the cross-sections at various lengths of an
example 7:1 NIR PLs during fabrication. The low panel shows the finished PL before cleaving and
splicing to standard fibers. }
\label{fig:photonic_lantern}
\end{figure}

A coronagraph-fed PIC spectrograph couples the planet light into a single-
mode fibre or a photonic lantern and disperses it on an
AWG~\citep{LeonSaval2005PL,Birks2015PL}. The single-mode input acts as a
spatial filter that rejects residual stellar speckle, improving the effective
contrast, a benefit unavailable to a conventional slit spectrograph. The
photonic lantern, an adiabatic taper that converts a multi-mode input into
several single-mode outputs, additionally allows partially-corrected light to
be accepted without loss of throughput, and the same device can serve as a
focal-plane wavefront sensor for low-order speckle
control~\citep{Norris2020GLINT}. For HWO, a coronagraph-fed AWG spectrograph at
$R\!\sim\!100$--$500$ over $0.2$--$1.8\,$\textmu m directly addresses the
Living Worlds biosignature driver. The low $R$ also removes the need for a cross-disperser typically used for separating the AWG's orders.
The photonic lantern that couples and reformats the coronagraphic beam into the single-mode AWG input is illustrated in Fig.~\ref{fig:photonic_lantern}; its mode-selective structure transitions adiabatically from a few-mode input to an array of single-mode outputs matched to the waveguide pitch of the AWG~\citep{LeonSaval2005PL,Birks2015PL}. The tapered structure accepts a partially corrected, few-mode beam from the coronagraph output and distributes it adiabatically into $N$ single-mode output fibres, each of which feeds one channel of the AWG.  The single-mode outputs simultaneously provide spatial filtering that suppresses residual stellar speckle before dispersal, improving effective coronagraphic contrast.  For HWO, a visible/NIR lantern covering $0.2$--$1.8\,\mu$m would serve the Living Worlds biosignature mode (driver~1 of Section~\ref{sect:drivers}).

\subsection{High-Resolution Cross-Correlation Spectroscopy}

HRCCS at $R\!\sim\!10^{5}$ is addressed by a tandem-AWG architecture
(Section~\ref{sect:awg}) in which a coarse AWG separates spectral orders and
a fine AWG provides the resolving power, with the dispersed output
butt-coupled to or cross-dispersed further before imaging on a detector array. Operating this architecture across the
UV-to-near-infrared band requires more than one material platform
(Section~\ref{sect:materials}).

\subsection{Multi-Object Spectroscopy}

\begin{figure}
\centering
\begin{subfigure}[b]{0.47\linewidth}
  \centering
  \includegraphics[width=\linewidth]{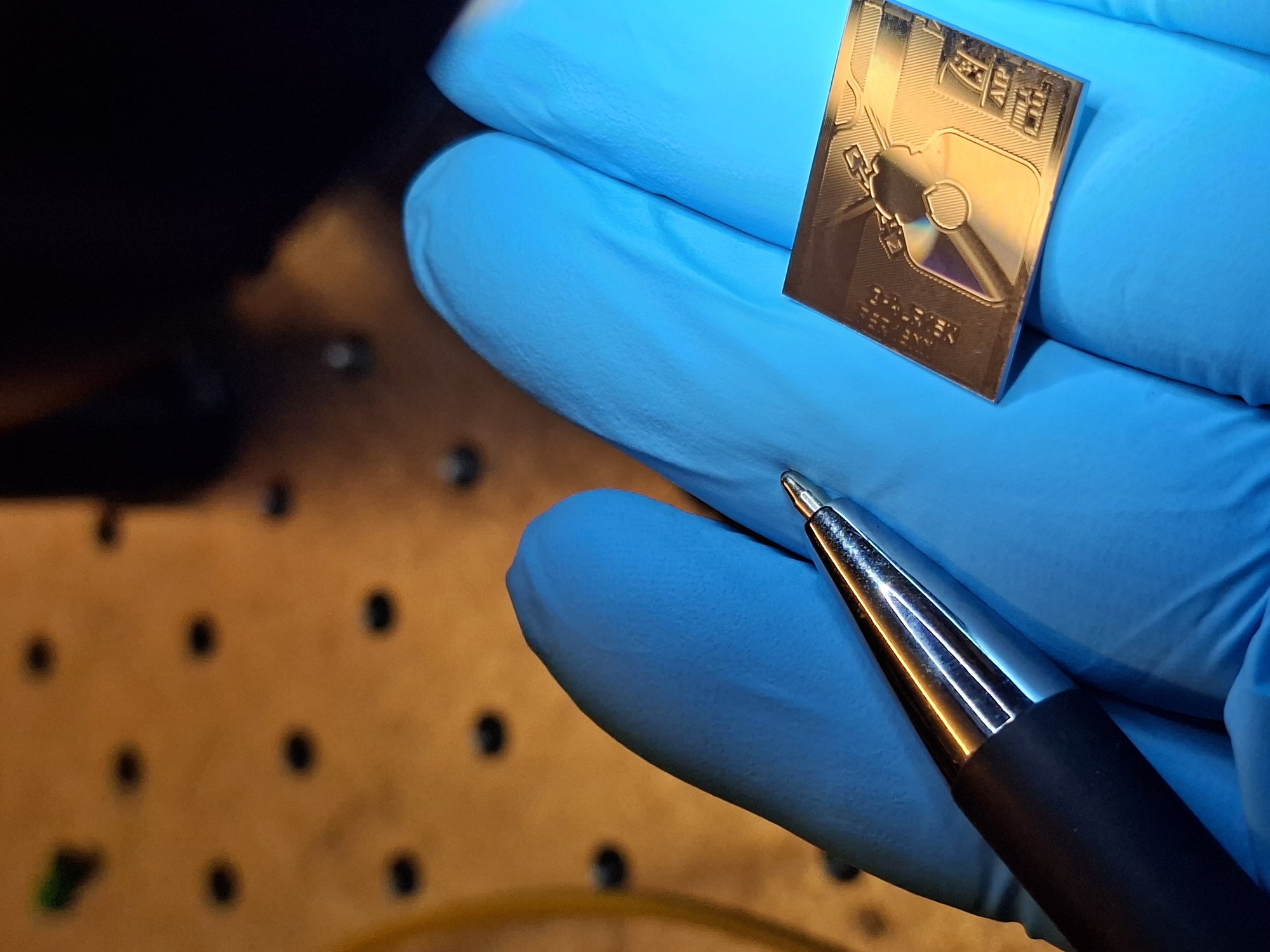}
  \caption{Astrophotonic AWG chip alongside a conventional grating.}
  \label{fig:size_awg}
\end{subfigure}
\hfill
\begin{subfigure}[b]{0.50\linewidth}
  \centering
  \includegraphics[width=\linewidth]{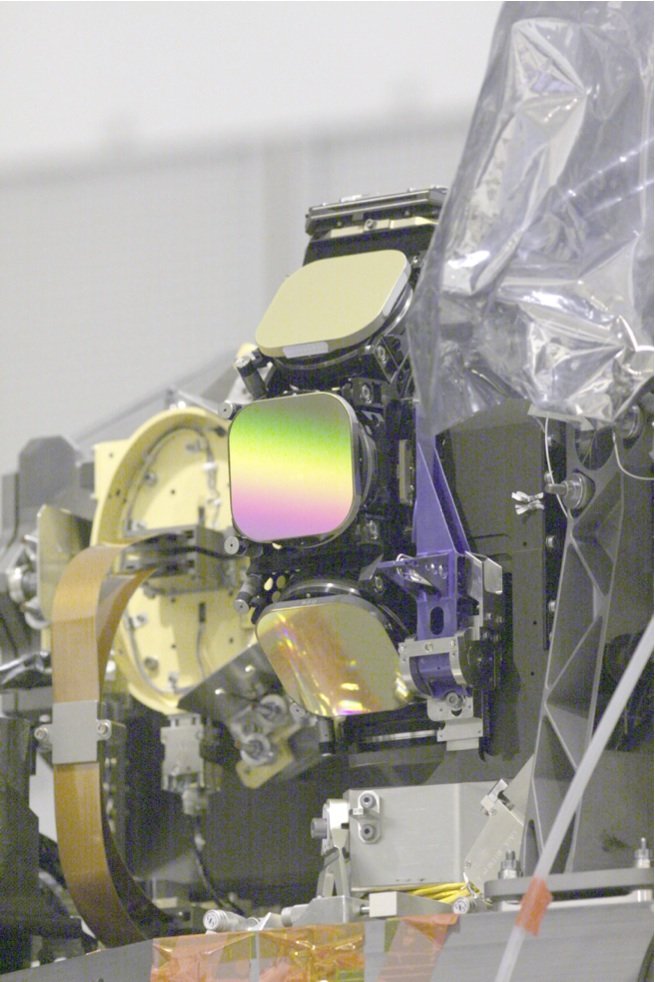}
  \caption{Grating Wheel Assembly (GWA) of JWST NIRSpec, manufactured by Carl
  Zeiss Optronics (Hensoldt). Credit: ESA/NASA. Reproduced under ESA's Open
  Access image policy; NASA imagery is not subject to copyright.}
  \label{fig:size_nirspec}
\end{subfigure}
\caption{Scale comparison: astrophotonic AWG chip versus the disperser subsystem of a state-of-the-art space spectrograph.
(\subref{fig:size_awg})~A silica-on-silicon AWG chip next to a conventional diffraction grating; the AWG chip fits comfortably in the palm of a hand and contains the complete dispersing function in a millimetre-scale guided-wave circuit.
(\subref{fig:size_nirspec})~The Grating Wheel Assembly (GWA) of JWST NIRSpec, showing six ruled gratings and two prisms on a cryogenic motor-driven mechanism~\citep{Jakobsen2022NIRSpec}. The GWA constitutes only the dispersing subsystem of NIRSpec.
For HWO, replacing a per-channel free-space optical train with a replicated guided-wave chip enables channel counts, $10^{3}$--$10^{4}$, within a chip footprint of order cm$^2$; this multiplexing density has not been demonstrated with conventional ruled-grating spectrographs~\citep{Cvetojevic2012AWG}.}
\label{fig:size_comparison}
\end{figure}

The defining advantage of PIC technology for multi-object spectroscopy~\citep{CAWSMOS,CAWSMOS2026} is
\emph{replication and, distribution}: once a single spectrograph channel is designed and
qualified, wafer-scale lithography produces hundreds or thousands of
identical or, distributed channels, at marginal incremental cost and mass. A microshutter or
fibre-positioner array selects targets at the focal plane; each target feeds
a selection of PIC spectrograph channels that can be identical or cover a wide wavelength spectrum. The mass and volume of the
resulting instrument scale far more slowly with multiplexing factor than for
any bulk-optic multi-object spectrograph~\citep{Cvetojevic2012AWG}, which is
decisive for a spectrograph with $10^{3}$--$10^{4}$ channels.
The scale of the challenge is placed in perspective by Fig.~\ref{fig:size_comparison}, which juxtaposes the AWG chip against the Grating Wheel Assembly (GWA) of the James Webb Space Telescope NIRSpec instrument, one of the most capable near-infrared multi-object spectrographs ever flown.
NIRSpec covers 0.6--5.3\,$\mu$m and provides multi-object spectroscopy via a microshutter assembly, but its disperser subsystem alone, six gratings and two prisms mounted on a cryogenic motor-driven wheel, illustrates the mechanical and volumetric complexity inherent to bulk-optic spectrograph design~\citep{Jakobsen2022NIRSpec}.
The NIRSpec gratings were precision-ruled and space-qualified by Carl Zeiss Optronics (now Hensoldt), and the full GWA mechanism including its cryogenic motor and position sensors occupies a volume and mass budget that is a significant fraction of the entire instrument~\citep{Jakobsen2022NIRSpec}.
The GWA is itself only the dispersing subsystem: NIRSpec additionally incorporates a fore-optics train, a filter wheel, a microshutter array with its control electronics, collimator and camera mirrors on a thermally stabilised optical bench, and a two-detector focal plane assembly, together representing an instrument with a mass of approximately 196\,kg and a volume envelope of $\sim$$1.9\times0.9\times0.7$\,m~\citep{Jakobsen2022NIRSpec}.
NIRSpec achieves a multiplexing factor of order 100 targets simultaneously.
A PIC multi-object spectrograph scales to $10^{3}$--$10^{4}$ channels with no analogous mechanical assembly: starlight enters a waveguide at the fibre-to-chip interface and propagates entirely in guided modes from input coupler through the AWG to the detector, with dispersion, order separation, and polarisation management all performed on-chip.
This \emph{fully guided} architecture eliminates the grating wheel mechanism, collimator, camera, bench alignment, and most of the thermal-stability hardware that instruments like NIRSpec require, and replaces the per-channel grating-plus-optics assembly with a lithographically replicated chip that can be produced at wafer scale with no moving parts and no free-space alignment interfaces.
For HWO, where launch mass, instrument volume, mechanism count, contamination risk, and long-term thermal and mechanical stability are all tightly constrained, this architectural difference has the potential to change a $10^{3}$--$10^{4}$-channel multi-object spectrograph from a prohibitively complex, mechanism-laden instrument into a viable one, provided that the space-qualification gaps identified in Sections~\ref{sect:space} and~\ref{sect:trl} are addressed.

\subsection{Integral-Field Spectroscopy}

A photonic integral-field unit uses a lenslet array or a multicore fibre to
sample the focal plane, a photonic-lantern or waveguide reformatter to map
the two-dimensional field onto a linear array of single-mode waveguides, and
an AWG to disperse the linearised output. Because the reformatting is
performed in guided optics, the instrument is compact and stable; the main
development need is low-loss UV operation (Section~\ref{sect:materials}).

\subsection{Extreme-Precision Radial Velocity Calibration}

EPRV imposes two photonic requirements. The spectrograph itself must be
wavelength-stable at the centimetre-per-second level, which favours a
thermally and mechanically stable PIC implementation; and the wavelength
calibrator must provide a dense, stable line grid. On-chip laser frequency
combs based on Kerr microresonators provide a gigahertz-spaced line grid over
hundreds of nanometres~\citep{Kippenberg2018Microresonator,Obrzud2019AstroComb},
and on-chip Fabry--P\'erot etalons provide a simpler, lower-cost
complement~\citep{Halverson2014FP}. Both are PIC components and both are
candidates for the HWO EPRV calibration subsystem.

\subsection{High-Resolution Spectropolarimetry}

Spectropolarimetry requires that the PIC preserve and analyse polarization
state. This requires polarization-maintaining waveguides, on-chip retarders,
and integrated polarization splitters. Birefringence-engineered AWGs and
thin-film lithium-niobate PICs~\citep{Pohl2020LNOI,Boes2018LNOI} are the
leading routes; the principal development need is control of waveguide
birefringence, discussed for AWGs in Section~\ref{sect:awg}.

\section{Material Platforms for UV-to-Near-Infrared PICs}
\label{sect:materials}

\subsection{The Wavelength-Coverage Challenge}

No single waveguide material spans the full HWO range from
$\sim\!100\,\mathrm{nm}$ to $2.5\,$\textmu m. Two physical constraints
dominate. First, every dielectric has a finite transparency window: silicon
and its oxide, the workhorses of telecommunications photonics, become
absorbing in the ultraviolet because the photon energy approaches or exceeds
the material bandgap. Second, a single-mode waveguide supports only about
half an octave of bandwidth before it either becomes multi-moded or guides
the light poorly. HWO photonic instruments will therefore necessarily be
built from \emph{several} material platforms, each optimised for a
sub-band, interfaced through hybrid integration.

\subsection{Waveguide Materials}

Table~\ref{tab:materials} summarises the leading waveguide platforms and
their approximate transparency windows. For ultraviolet and blue
wavelengths, wide-bandgap materials are required: aluminium nitride is
transparent to wavelengths as short as $\sim\!200\,\mathrm{nm}$, amorphous
aluminium oxide to $\sim\!225\,\mathrm{nm}$ (with sub-dB/cm propagation
losses demonstrated at $360\,\mathrm{nm}$ and active CMOS-fabricated devices
operating at $320\,\mathrm{nm}$), and tantalum pentoxide to
$\sim\!300\,\mathrm{nm}$~\citep{Blumenthal2020,Castillo2026AluminaUV}. Silicon nitride is transparent
across a remarkably broad band (approximately $400\,\mathrm{nm}$ to
$2.4\,$\textmu m) and has demonstrated ultra-low propagation loss, which
makes it the natural workhorse for the optical and near-infrared. Silica
remains attractive in the visible and near-infrared because of its excellent
fibre-coupling efficiency and low loss. Thin-film lithium niobate adds
electro-optic functionality useful for active phase control and
polarization handling. Beyond $2.5\,$\textmu m, chalcogenide glasses and
silicon/germanium platforms become necessary, although they fall outside the
HWO band and are listed only for completeness.

\begin{table*}
\caption{Leading PIC waveguide material platforms, approximate transparency
windows, and relevance to the HWO wavelength range. Transparency limits are
approximate and depend on film quality and acceptable loss.}
\label{tab:materials}
\begin{center}
\footnotesize
\begin{tabularx}{\textwidth}{|l|l|X|}
\hline
\rule[-1ex]{0pt}{3.5ex}
\textbf{Platform} & \textbf{Transparency (approx.)} & \textbf{Role for HWO} \\
\hline\hline
\rule[-1ex]{0pt}{3.5ex}
Aluminium nitride (AlN) & $\gtrsim\!200\,\mathrm{nm}$ to NIR &
Far/near-UV waveguides and AWGs \\
\hline
\rule[-1ex]{0pt}{3.5ex}
Aluminium oxide (Al$_2$O$_3$, ALD) & $\gtrsim\!225\,\mathrm{nm}$ to NIR &
Near-UV waveguides and active PICs; CMOS filter at 320\,nm demonstrated~\protect\citep{Castillo2026AluminaUV}; AWG not yet demonstrated \\
\hline
\rule[-1ex]{0pt}{3.5ex}
Tantalum pentoxide (Ta$_2$O$_5$) & $\gtrsim\!300\,\mathrm{nm}$ to NIR &
Near-UV to optical waveguides \\
\hline
\rule[-1ex]{0pt}{3.5ex}
Silicon nitride (Si$_3$N$_4$) & $\sim\!0.4$--$2.4\,$\textmu m &
Optical/NIR workhorse; low-loss high-$R$ AWGs \\
\hline
\rule[-1ex]{0pt}{3.5ex}
Silica (SiO$_2$, doped) & $\sim\!0.35$--$2.3\,$\textmu m &
Optical/NIR AWGs; excellent fibre coupling \\
\hline
\rule[-1ex]{0pt}{3.5ex}
Lithium niobate (LiNbO$_3$, thin film) & $\sim\!0.4$--$5\,$\textmu m &
Active phase/polarization control; electro-optic devices \\
\hline
\rule[-1ex]{0pt}{3.5ex}
Chalcogenide / Si-Ge & $\gtrsim\!2\,$\textmu m (mid-IR) &
Beyond the HWO band; listed for completeness \\
\hline
\end{tabularx}
\end{center}
\end{table*}

The ultraviolet remains the least mature regime: UV PIC development trails
the near-infrared because Rayleigh scattering, which rises steeply toward
short wavelengths, places stringent demands on waveguide sidewall roughness,
and because wide-bandgap thin films of suitable purity are harder to grow and
pattern. Recent demonstrations of UV ring resonators and UV arrayed waveguide
gratings~\citep{Blumenthal2020} establish feasibility, but bringing UV PICs to
the maturity of their near-infrared counterparts is the single
most-leveraged material-development investment for HWO, since UV access is
required by forty-one percent of the SCDDs.

\subsection{Optical Fibre Materials}

Optical fibres feed light from the telescope focal plane to the PIC and
between subsystems. Standard UV-grade fused silica transmits into the
near-ultraviolet but suffers solarization (radiation- and UV-induced
absorption from colour-centre formation), which limits reliable operation
at the shortest wavelengths~\citep{Girard2019}. Fluoride and ZBLAN glasses
extend transmission but are mechanically brittle and limited to piece lengths
of order one hundred metres. Anti-resonant hollow-core fibres, in which light
is guided in an air core, largely circumvent material absorption and
solarization; near-infrared anti-resonant hollow-core fibres have reached
losses of $0.174\,\mathrm{dB\,km^{-1}}$~\citep{Jasion2022HCF}, and the
technology is being extended toward the ultraviolet and mid-infrared.
Multicore silica fibres support compact integral-field feeds. For HWO, a
combination of UV-grade silica and solarization-resistant hollow-core fibres
is the most plausible fibre solution across the band. Table~\ref{tab:fibresurvey} summarises the availability of optical fibres covering the HWO wavelength range ($\sim$0.1--2.5\,\textmu m), with space-qualification readiness and representative vendor products.
\begin{itemize}
  \item Space environment and dose : A shielded satellite interior receives
on average $\sim$0.05\,Rad/min; over a 5--10\,year mission this accumulates
to $\sim$125--250\,kRad of total ionising dose
\citep[][a low-Earth-orbit estimate]{Alam}. HWO will operate at the Sun--Earth L2 point, where the
particle mix differs, with fewer trapped protons, more galactic cosmic rays and
solar-event particles, but a total ionising dose of order a few kRad to a
few hundred kRad behind shielding remains a reasonable planning figure.
Optical fibres darken under dose through colour-centre formation
(radiation-induced attenuation, RIA); fibres with a pure-silica core and a
fluorine-doped cladding are the most radiation-hard, whereas Ge-, P-, Al-,
and rare-earth dopants worsen RIA.
  \item Termination is part of qualification: A bare fibre becomes a flight
assembly only once terminated and cabled. Diamond SA supplies
ESA-ESCC-qualified single-channel fibre-optic terminations (Mini-AVIM,
ESCC 3420/001; an ESA-qualified supplier since 2018) compatible with SM, MM,
PM, and multicore fibre, with flight heritage including NASA's Perseverance
rover. Space-qualified fibre and space-qualified termination are
complementary requirements.
\item ``Space-qualification readiness''
distinguishes formally space-qualified or flight-proven products from those
only laboratory-irradiation-tested and from research-grade fibres; it is a
qualitative indicator, not a formal Technology Readiness Level. Diameters,
NA, wavelength limits, and RIA figures are representative values that vary
between vendors, batches, and irradiation conditions.

\end{itemize}

\clearpage
\onecolumn
\begin{landscape}
\begingroup
\setlength{\textwidth}{9.3in}\setlength{\linewidth}{9.3in}
\setlength{\LTcapwidth}{9.3in}
\setlength{\tabcolsep}{3pt}
\centering
\scriptsize
\renewcommand{\arraystretch}{1.2}

\begin{longtable}{%
  |>{\raggedright\arraybackslash}p{2.0cm}%
  |>{\raggedright\arraybackslash}p{1.7cm}%
  |>{\raggedright\arraybackslash}p{1.55cm}%
  |>{\centering\arraybackslash}p{0.8cm}%
  |>{\centering\arraybackslash}p{0.8cm}%
  |>{\raggedright\arraybackslash}p{0.85cm}%
  |>{\raggedright\arraybackslash}p{4.2cm}%
  |>{\raggedright\arraybackslash}p{4.5cm}%
  |>{\raggedright\arraybackslash}p{3.6cm}|}

\caption{Optical fibres for the Habitable Worlds Observatory wavelength range
($\sim$0.1--2.5\,\textmu m), with space-qualification readiness. For each
fibre family~\protect\citep{Alam,Girard_2018,Colombe14,Jasion2022HCF}: transmission / single-mode wavelength range, representative
core/cladding diameters, availability as single-mode (SMF) and/or multimode
(MMF) fibre, numerical aperture (NA), representative products and vendors,
space-qualification readiness, and whether the fibre has been used to
fabricate photonic lanterns. \Y\ = yes; \N\ = no / not applicable.}
\label{tab:fibresurvey}\\
\hline
\rowcolor[gray]{0.80}
\textbf{Fibre family} & \textbf{Wavelength range} &
\textbf{Core / clad (\textmu m)} & \textbf{SMF} & \textbf{MMF} &
\textbf{NA} & \textbf{Representative products / vendors} &
\textbf{Space-qualification readiness} &
\textbf{Photonic lanterns?}\\
\hline
\endfirsthead
\multicolumn{9}{l}{\footnotesize\itshape Table~\thetable\ (continued)}\\[2pt]
\hline
\rowcolor[gray]{0.80}
\textbf{Fibre family} & \textbf{Wavelength range} &
\textbf{Core / clad (\textmu m)} & \textbf{SMF} & \textbf{MMF} &
\textbf{NA} & \textbf{Representative products / vendors} &
\textbf{Space-qualification readiness} &
\textbf{Photonic lanterns?}\\
\hline
\endhead
\hline
\multicolumn{9}{r}{\footnotesize\itshape continued on next page}\\
\endfoot
\hline
\endlastfoot

$\sim$0.18--1.2\,\textmu m; solarises $<$0.30\,\textmu m &
50--1000 / 125--1100 &
\N & \Y & 0.22 &
Thorlabs FG-series; Polymicro FV-series; \textbf{j-fiber} UV /
solarisation-resistant MM grades &
Pure-silica core gives useful radiation tolerance; laboratory $\gamma$ and
proton data exist. Deep-UV solarisation limits life (treat as a consumable).
No standard flight qualification. &
Indirectly, silica MMF of this class is the multimode input of
visible/NIR lanterns; no UV lantern demonstrated.\\
\hline

Solarisation-resistant UV MMF (H$_2$-treated, Al-coated) &
$\sim$0.19--2.4\,\textmu m &
50--400 / 125--440 &
\N & \Y & 0.12--0.275 &
Fiberguide Solarguide UVS-H2A; Thorlabs FG200UVA; \textbf{j-fiber}
solarisation-resistant MM &
Suborbital UV heritage (e.g.\ FIREBall balloon); lab-tested. Deep-UV
solarisation still life-limiting. Not formally space-qualified. &
No, no UV photonic lantern demonstrated to date.\\
\hline

Single-mode $\sim$0.28--0.40\,\textmu m (research) &
large-mode-area core / $\sim$125 &
\Y & \N & low ($\sim$0.05--0.10) &
NIST H$_2$-loaded PCF (Colombe et al.\ 2014); NKT LMA-UV &
Research-grade; not space-qualified. &
No, UV single-mode fibre still at the research stage.\\
\hline

Visible single-mode step-index fibre &
Single-mode $\sim$0.40--0.68\,\textmu m &
$\sim$2--3 / 125 &
\Y & \N & $\sim$0.12--0.14 &
Thorlabs S405-XP, SM400, 460HP; radiation-tolerant visible grades from
\textbf{Coherent} (\textbf{Nufern}) and \textbf{Fibercore} &
COTS telecom-type; standard grades are not radiation-hard.
Pure-silica-core / radiation-tolerant variants reduce radiation-induced
attenuation (RIA), verify per fibre. &
Yes, visible photonic lanterns demonstrated from this class of fibre.\\
\hline

Single-mode $\sim$0.98--1.65\,\textmu m &
4.5--8.2 / 125 (or 80) &
\Y & \N & 0.11--0.16 &
Corning SMF-28 Ultra; \textbf{Fibercore} SM980; \textbf{Coherent}/%
\textbf{Nufern} S1550-HTA (rad-hard), R1310-HTA (rad-tolerant);
\textbf{Exail} IXF-RAD-SM-1550-0.14-PI; \textbf{j-fiber}
radiation-resistant SM &
\textbf{Best-developed regime.} Space-qualified radiation-hard /
-tolerant SM fibre available: pure-silica-core SM (S1550-HTA) RIA
$\sim$1\,dB/km at 50\,kRad ($\gamma$), saturating early; rad-tolerant
SMF-28-type $\sim$30\,dB/km at 1\,MRad. \textbf{Exail} rad-hard SM has
flight heritage on tens of satellites. &
Yes, the workhorse fibre; 19-port silica lanterns reach $>$95\,\%
throughput at 1.55\,\textmu m.\\
\hline

Polarization-maintaining (PM) fibre &
$\sim$0.83--1.65\,\textmu m (per fibre) &
$\sim$4.5--7 / 80 or 125 &
\Y & \N & $\sim$0.12--0.20 &
\textbf{Coherent}/\textbf{Nufern} PM1550G-80 \& PM850G-80 (Panda,
gyro-grade); \textbf{Exail} IXF-PMG space-grade gyro fibres;
\textbf{Fibercore} Radiation-Tolerant PM (bow-tie gyro fibre) &
Strong flight heritage through fibre-optic gyroscopes; gyro-grade
radiation-tolerant PM fibre flown extensively. $\gamma$-RIA
$\sim$6\,dB/m at 50\,kRad (1.55\,\textmu m); 0.83\,\textmu m PM worse
($\sim$150\,dB/m). &
Not typically, PM fibre is used for polarisation-preserving feeds and
components; relevant to HWO spectropolarimetry.\\
\hline

Vis--NIR, $\sim$0.5--2.3\,\textmu m &
$N$ cores in one 125+ cladding &
\Y & \N & $\sim$0.1--0.2 per core &
Custom draws (OFS, \textbf{Fibercore}, university facilities) &
No standard space qualification; inherits silica radiation behaviour;
custom development required. &
Yes, multicore silica fibre forms the single-mode output array of
standard photonic lanterns.\\
\hline

Anti-resonant hollow-core fibre (air-guiding silica microstructure) &
Design-dependent across UV--NIR; 0.174\,dB/km at 1.55\,\textmu m &
hollow core 10--50 / 125--250 &
\Y\textsuperscript{a} & \Y & very low (air) &
GLOphotonics; Lumenisity NANF; Jasion et al.\ 2022 (DNANF) &
Not yet space-qualified; the air core (little glass in the light path)
makes it intrinsically radiation-tolerant, promising but at R\&D
stage. &
R\&D, hollow-core / anti-resonant lanterns explored, not yet standard.\\
\hline

Transmits 0.285--4.5\,\textmu m; single-mode 2.3--4.1\,\textmu m &
6.5 / 125 (SM); 7.5--680 core (MM) &
\Y & \Y & $\sim$0.20 &
Thorlabs ZFG single-mode; Le Verre Fluor\'e; art photonics &
Not space-qualified; fluoride-glass radiation response is less
characterised than silica; hygroscopic and mechanically fragile, a
significant qualification gap. &
Partial, single-mode ZBLAN fibre couplers demonstrated (lantern
building block).\\
\hline

Fluoride fibre, InF$_3$ (fluoroindate) &
Transmits 0.31--5.5\,\textmu m; single-mode 3.2--5.5\,\textmu m &
7.5--9 / 125 (SM); to 600+ core (MM) &
\Y & \Y & $\sim$0.2--0.3 &
Thorlabs InF$_3$ single-mode; art photonics; Le Verre Fluor\'e &
Not space-qualified; radiation response less characterised than silica;
hygroscopic and mechanically fragile, as ZBLAN. &
Yes (recent), the first mid-infrared fibre photonic lantern was
demonstrated using InF$_3$ fibres.\\
\hline

\end{longtable}

\vspace{0.6em}
\begin{minipage}{9.3in}
\scriptsize
\textsuperscript{a}\,Anti-resonant hollow-core fibres are effectively
single-mode (low higher-order-mode content) by design rather than by a
conventional core/cladding index step.\\[3pt]
\end{minipage}

\endgroup
\end{landscape}
\clearpage
\twocolumn

\section{Photonic Spectrographs Using Arrayed Waveguide Gratings}
\label{sect:awg}

\subsection{Operating Principle and Astronomical Requirements}

\begin{figure}
\centering
\begin{subfigure}[b]{0.42\linewidth}
  \includegraphics[width=\linewidth]{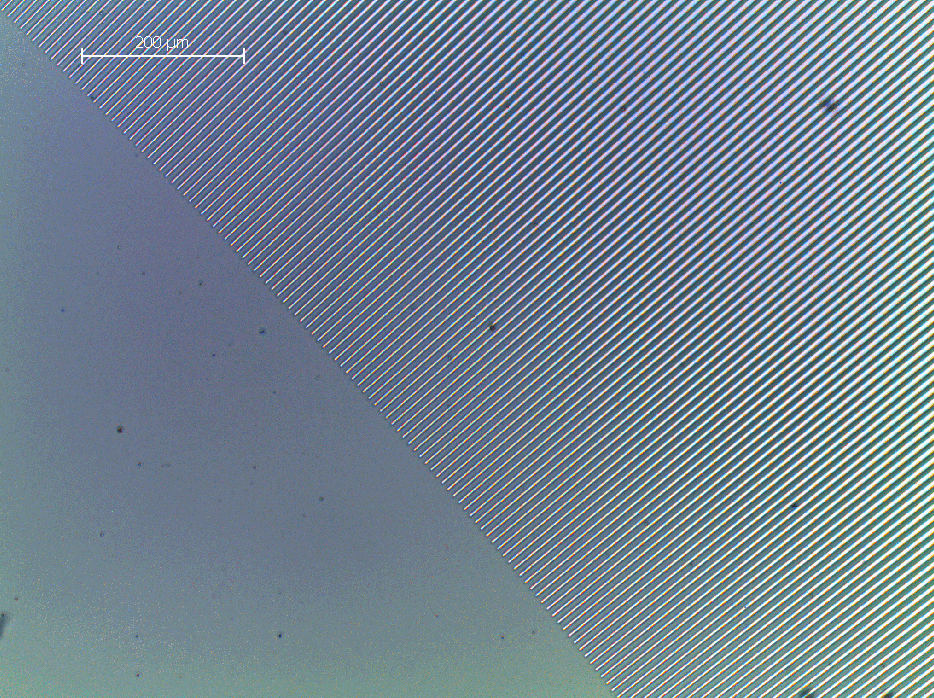}
  \caption{Microscope image of the AWG waveguide grating array (1\,$\times$).}
  \label{fig:awg_chip_1x}
\end{subfigure}
\hfill
\begin{subfigure}[b]{0.42\linewidth}
  \includegraphics[width=\linewidth]{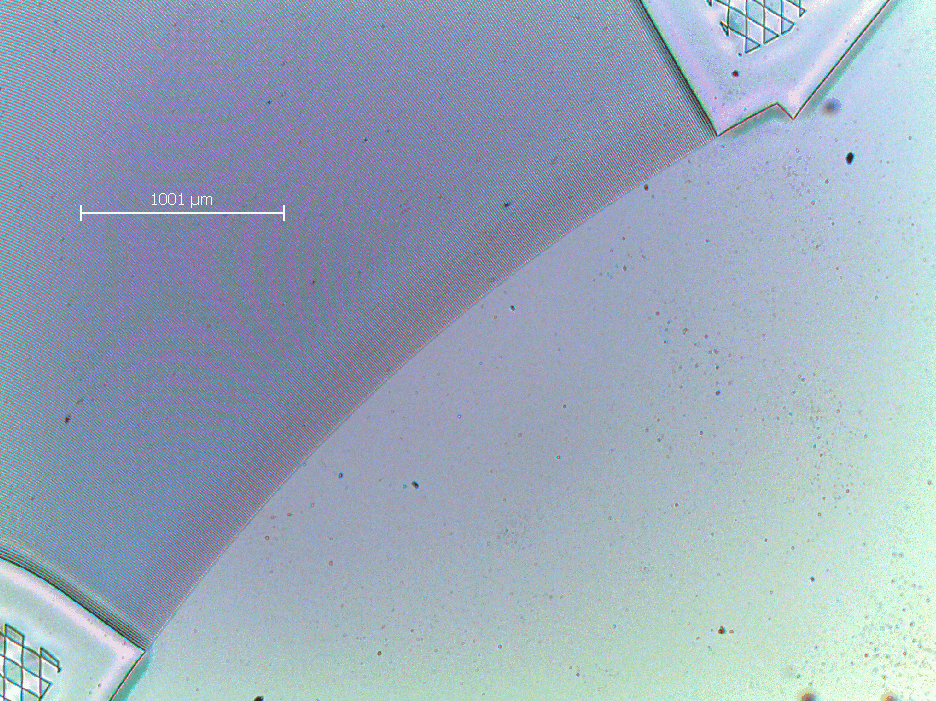}
  \caption{Microscope image of the waveguide array (2.5\,$\times$ zoom) looking into the free propagation region.}
  \label{fig:awg_chip_2p5x}
\end{subfigure}
\caption{Microscope images of the silica-on-silicon AWG waveguide grating array fabricated for the PAWS programme.  (\subref{fig:awg_chip_1x})~Overview of the arrayed waveguide region showing the individual waveguide arms at 1$\times$ magnification.  (\subref{fig:awg_chip_2p5x})~2.5$\times$ magnified view revealing the waveguide pitch, sidewall quality, and inter-waveguide spacing of the grating array.  The lithographic uniformity of the waveguide array visible in these images sets the phase-error floor discussed in Section~\ref{subsect:phaseerror}.}
\label{fig:awg_chips}
\end{figure}

An arrayed waveguide grating disperses light by splitting an input into an
array of waveguides of incrementally increasing path length; the resulting
wavelength-dependent phase tilt focuses different wavelengths to different
points on an output free-propagation region. AWGs are a mature
telecommunications technology, but astronomical use imposes more demanding
specifications: high throughput, broad bandwidth, high resolving power
($R\!\equiv\!\lambda/\Delta\lambda\!\gtrsim\!10{,}000$), and polarization
independence~\citep{Gatkine2019Review,Blind2017Spectrographs}. AWGs are
preferred over the alternative photonic echelle gratings because they avoid
the difficult fabrication of smooth on-chip reflecting facets while still
delivering efficiencies above seventy-five percent. Astrophotonic AWGs on a
two-percent-index-contrast silica platform have demonstrated with resolving powers
of $R\!\sim\!10{,}000$--$36{,}000$ with efficiencies up to seventy-two
percent in the near-infrared $H$ band~\citep{Stoll21}. Theoretically, these devices can reach $R\!\sim\!650{,}000$~\citep{Stoll20}.  The Potsdam Arrayed Waveguide
Spectrograph (PAWS) and the Compact Arrayed Waveguide Stacked Multi-Object Spectrograph (CAWSMOS) are active development platforms in this
class~\citep{Stoll2020PAWS,Stoll2017PAWS,CAWSMOS,CAWSMOS2026}.
Representative AWG chips from the PAWS programme are shown in Fig.~\ref{fig:awg_chips}, illustrating the physical scale and optical architecture of the device~\citep{Stoll2020PAWS}.

\begin{figure}
\centering
\includegraphics[width=0.62\linewidth]{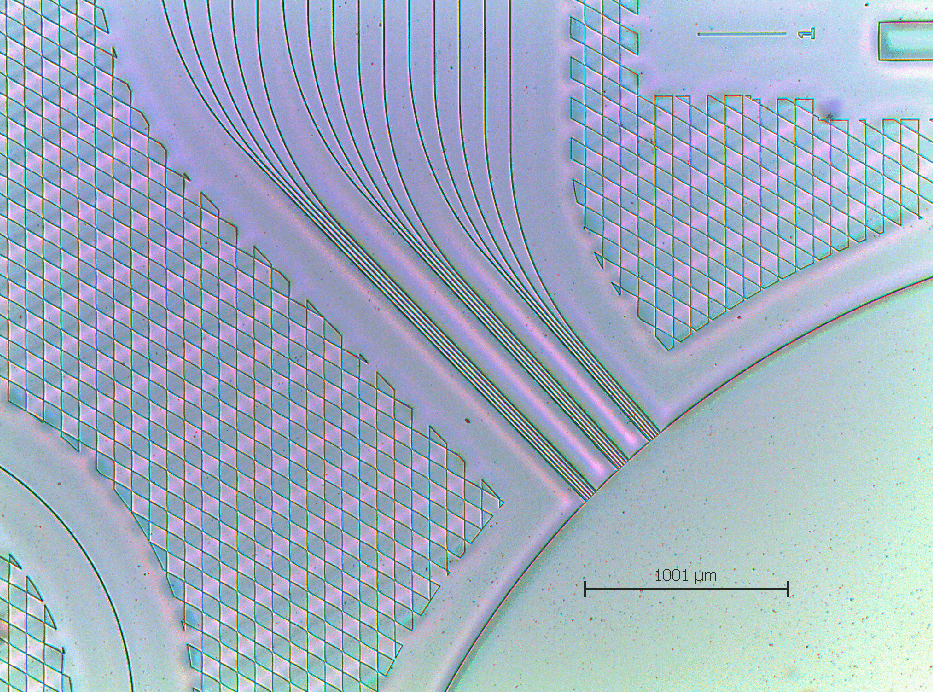}
\caption{Microscope image looking along the 15 input waveguides of an AWG toward the entrance of the free-propagation region (FPR).  The regular spacing and the smooth waveguide facets are critical: any lithographic roughness or pitch non-uniformity introduces phase errors across the array, which broaden the point-spread function on the output focal surface and limit the achievable resolving power.  For HWO-class resolving powers ($R\!\gtrsim\!10^{4}$), path-length uniformity of better than ten parts per million is required across the full waveguide array, motivating the tight process control and post-fabrication trimming strategies described in Section~\ref{subsect:phaseerror}.}
\label{fig:awg_waveguides_fpr}
\end{figure}

\subsection{Reaching High Resolving Power: the Phase-Error Challenge}
\label{subsect:phaseerror}

\begin{figure}
\centering
\begin{subfigure}[b]{0.55\linewidth}
  \includegraphics[angle=-180, width=\linewidth]{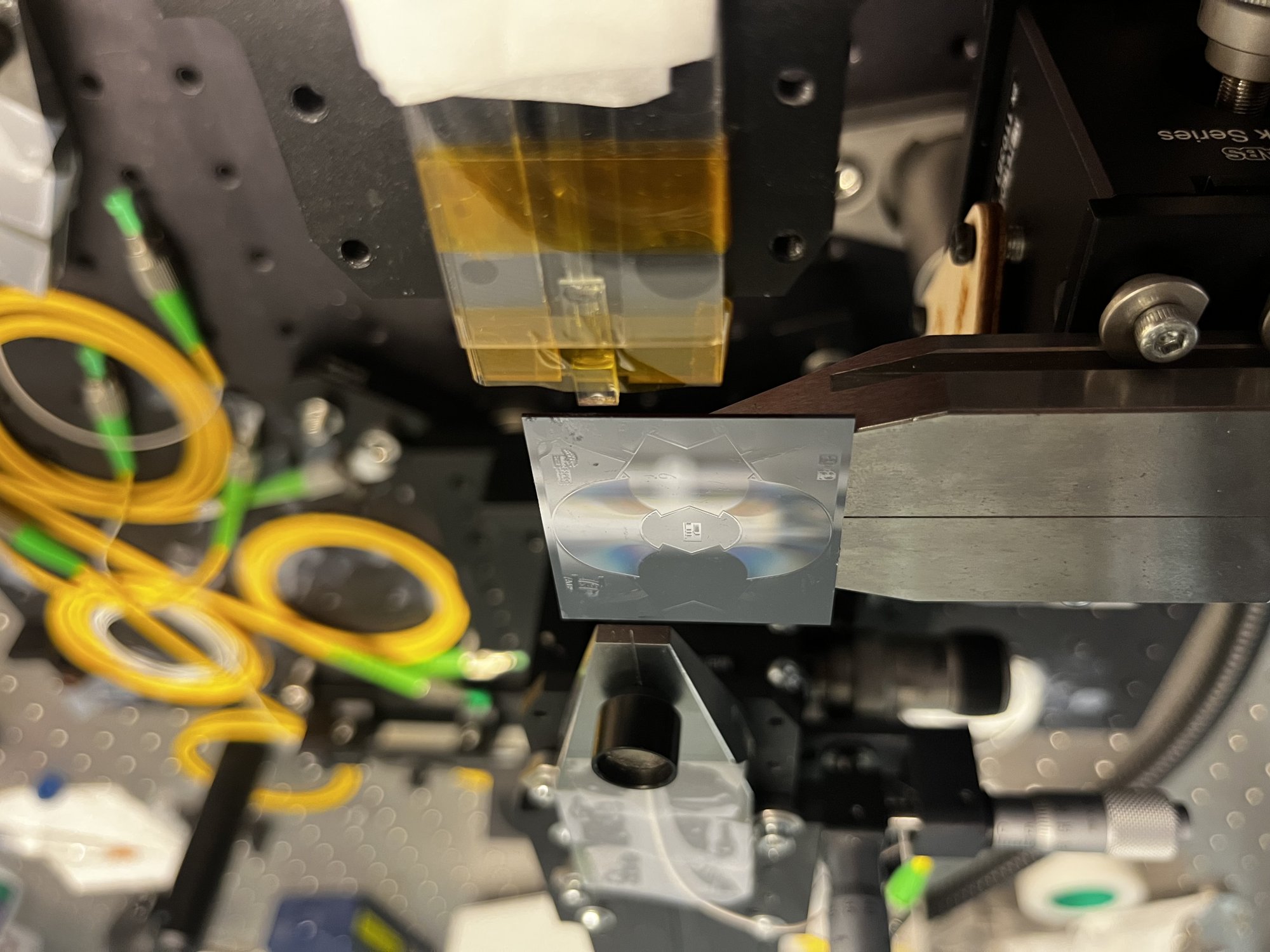}
  \caption{Single undiced AWG with 16-fibre array.}
  \label{fig:undiced_single}
\end{subfigure}
\hfill
\begin{subfigure}[b]{0.40\linewidth}
  \includegraphics[width=\linewidth]{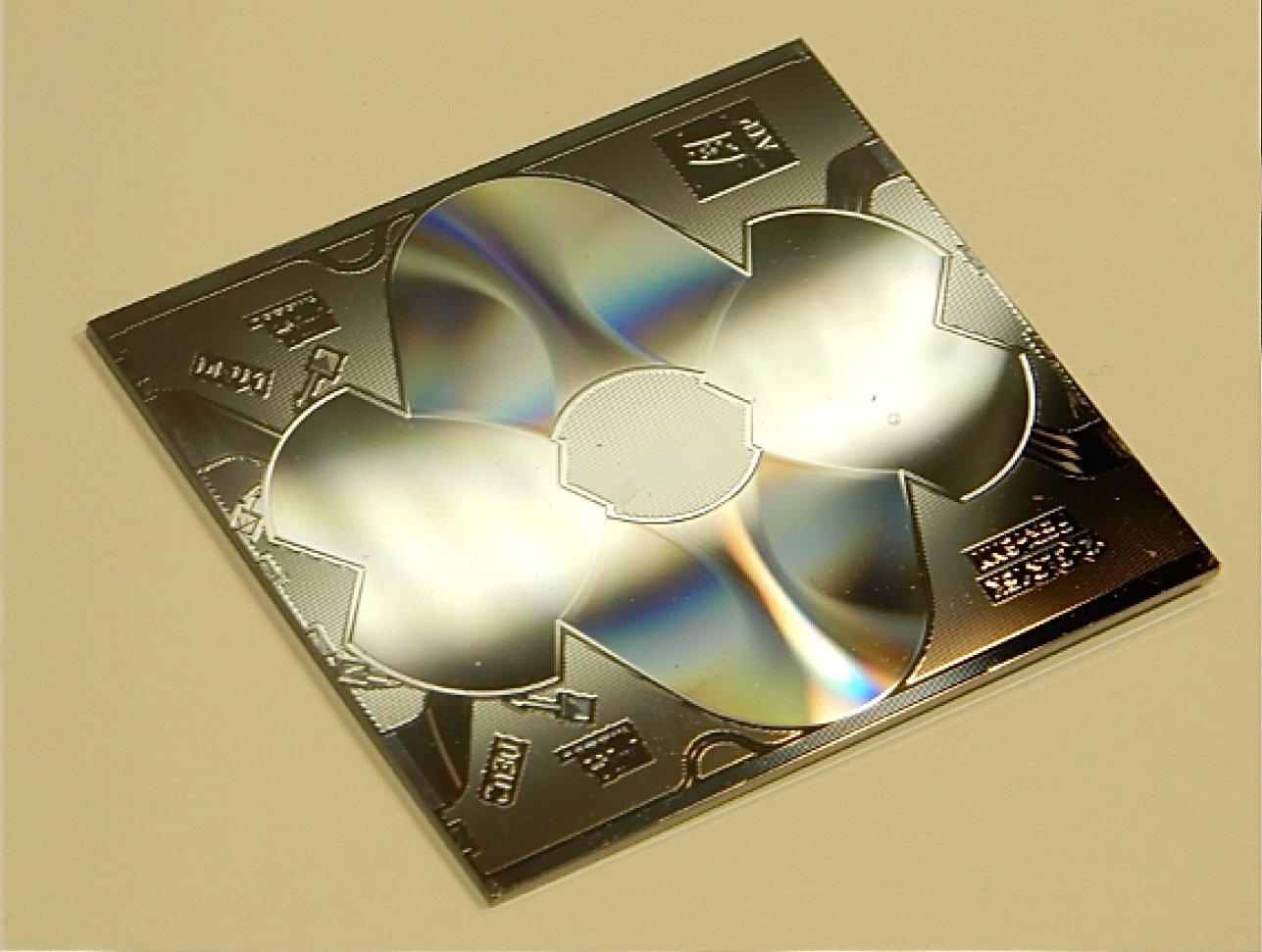}
  \caption{Photograph of an undiced double-AWG chip; each AWG is accessible from 15 fibres and the two devices can differ in resolving power $R$ and free spectral range (FSR).}
  \label{fig:undiced_double}
\end{subfigure}
\caption{Photographs of AWG chips at the pre-dicing, wafer-level characterisation stage.
(\subref{fig:undiced_single})~A single undiced silica-on-silicon AWG chip with a 16-element fibre ribbon array butt-coupled to the input facets for laboratory insertion-loss and phase-error measurement prior to dicing.  This approach allows defective devices to be rejected early in the flow, reducing fabrication waste and characterising the phase-error statistics of a production batch.
(\subref{fig:undiced_double})~Photograph of an undiced double-AWG chip, in which two AWG devices sharing a common wafer footprint are each accessible from 15 fibres and can differ in resolving power $R$ or free spectral range (FSR).  Such a dual-design layout allows direct, same-wafer comparison of architectural choices (e.g.\ Rowland versus anastigmatic geometry, or two FSR values) under identical fabrication conditions, directly informing the cascaded-AWG cross-dispersion strategy of Section~\ref{subsect:crossdisp}.}
\label{fig:undiced_awgs}
\end{figure}

High resolving power requires a large number of long arrayed waveguides, and
therefore a large device footprint. Across a large chip, material and process
variations produce systematic and random variations in the waveguide
effective index, which translate into optical-path-length errors. These phase
errors reduce the power in the main diffraction lobe and raise crosstalk, a fundamental performance limit studied extensively in the telecommunications
AWG literature~\citep{Smit1996AWGReview,Yamada1998PhaseError}.
An AWG at $R\!\gtrsim\!24{,}000$ requires control of the optical path lengths to
of order ten parts per million, which becomes increasingly difficult as the
waveguide count exceeds a few hundred and the accumulated path length reaches
several centimetres. Three mitigation strategies exist: post-fabrication
trimming or active phase shifters (electro-optic, thermo-optic, or
piezoelectric) on each waveguide~\citep{Yamada1998PhaseError}; minimal-phase-error
designs and tighter process control that eliminate the errors at the source;
and novel compact architectures such as the reusable-delay-line AWG, which can
be roughly two orders of magnitude smaller than a conventional AWG and so is
intrinsically less susceptible to chip-scale process variation~\citep{rdl}.
Higher-index-contrast platforms such as silicon nitride allow much smaller
bend radii and hence smaller, less variation-prone devices~\citep{Smit1996AWGReview}.
The wafer-level characterisation approach that underpins phase-error diagnosis is illustrated in Fig.~\ref{fig:undiced_awgs}: chips are evaluated with a fibre array while still on the parent wafer, enabling systematic mapping of phase-error statistics across a production run before the costly dicing and facet-polishing steps that finalise individual devices.

\subsection{Order Separation and Cross-Dispersion}
\label{subsect:crossdisp}

\begin{figure}
\centering
\includegraphics[width=0.62\linewidth]{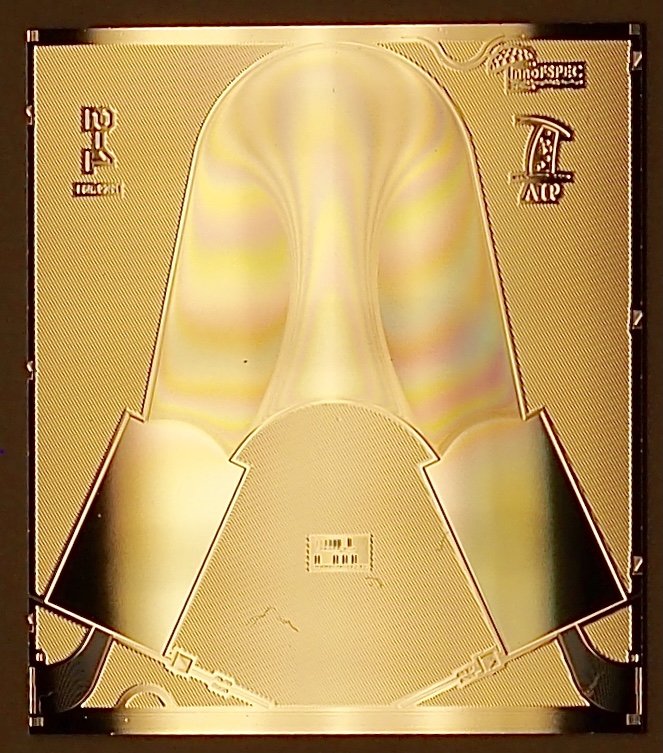}
\caption{Photograph of a three-stigmatic-point (anastigmatic) silica-on-silicon AWG chip from the PAWS programme.}
\label{fig:stigmatic_awg}
\end{figure}

A high-resolution AWG has a free spectral range of only $10$--$20\,\mathrm{nm}$,
so covering an astronomical band requires order separation. Semi-integrated
spectrographs use bulk-optic cross-dispersion, which limits miniaturisation. The corresponding echellogram recorded with the full PAWS instrument is shown in Section~\ref{subsect:paws} (Fig.~\ref{fig:paws_frame}). 
A fully integrated alternative is the tandem (cascaded) AWG, in which a coarse
AWG separates orders and a fine AWG provides resolving power, allowing the
dispersed output to be edge-coupled directly to a detector array. Realising
low-loss tandem AWGs requires ultra-low-loss single-stage AWGs and
minimisation of inter-channel spectral dropout in the coarse stage. A flat
focal plane, achievable with a three-stigmatic-point design, further allows
the detector to be bonded directly to the chip edge.
The stigmatic design is illustrated in Fig.~\ref{fig:stigmatic_awg}, which shows a photograph of a three-stigmatic-point silica AWG chip developed in the PAWS programme~\citep{Stoll2020PAWS}. The anastigmatic layout achieves diffraction-limited focus at three wavelengths symmetrically distributed across the free spectral range, flattening the focal surface across the full spectral order and allowing a planar detector (or detector-bonded chip edge) to be used without refocusing optics.  This design is also directly applicable to the cross-disperser architecture required for HWO's high-$R$ modes, where the  output must lie on a flat focal surface to couple efficiently into the relay optics and detector~\citep{Stoll2020PAWS,rdl}.

\subsection{Throughput Budget}

Table~\ref{tab:throughput} gives a representative end-to-end throughput budget for a silica/SiN AWG spectrograph, now extended to include detector quantum efficiency so that the final entry is a true photon-to-electron conversion efficiency~\citep{Jovanovic2023Roadmap}.
Without detector QE, the optical efficiency from fibre input to free-space output (``optical-only'') improves from $\sim$27\,\% (unoptimised) to $\sim$79\,\% (optimised); including a realistic detector QE of 70\,\% (current) or 85\,\% (optimised HgCdTe / back-illuminated CMOS), the true photon-to-electron end-to-end efficiency is $\sim$19\,\% and $\sim$67\,\% respectively for the two cases.
The optimised value assumes a directly coupled detector (Section~\ref{subsect:detector_int}), in which the relay optics loss is eliminated, together with a single-order AWG design that removes the cross-dispersion loss term entirely.
Because exposure time scales inversely with throughput at fixed signal-to-noise ratio, this improvement, a factor of nearly four relative to the current unoptimised baseline, is decisive for photon-starved HWO observations.

\begin{table}
\caption{Representative end-to-end throughput budget for a silica AWG spectrograph from fibre input to detector output: current state of the art and the value achievable with optimised components.
Detector quantum efficiency (QE) is included to give a true photon-to-electron conversion efficiency; values quoted are for the near-infrared $H$-band ($\sim$1.5\,$\mu$m).
Where the detector is attached directly at the AWG output facet (Section~\ref{subsect:detector_int}), the relay optics loss is eliminated.}
\label{tab:throughput}
\begin{center}
\footnotesize
\begin{tabularx}{\linewidth}{|X|c|c|}
\hline
\rule[-1ex]{0pt}{3.5ex}
\textbf{Component} & \textbf{Current} & \textbf{Optimised} \\
\hline\hline
\rule[-1ex]{0pt}{3.5ex}
Single-mode-fibre-to-chip coupling & 95\,\% & 95\,\% \\
\hline
\rule[-1ex]{0pt}{3.5ex}
AWG throughput (silica / SiN) & 70\,\% & 85\,\% (optimised tapers) \\
\hline
\rule[-1ex]{0pt}{3.5ex}
AWG output coupling / relay optics & 80\,\% & 98\,\% (direct-coupled detector) \\
\hline
\rule[-1ex]{0pt}{3.5ex}
Cross-dispersion (if required) & 50\,\% & 100\,\% (single-order design) \\
\hline
\rule[-1ex]{0pt}{3.5ex}
Detector quantum efficiency & 70\,\% (CMOS / HgCdTe) & 85\,\% (back-illuminated) \\
\hline
\rule[-1ex]{0pt}{3.5ex}
Optical-only subtotal (excluding detector QE) & 27\,\% & 79\,\% \\
\hline
\rule[-1ex]{0pt}{3.5ex}
\textbf{End-to-end (photon $\to$ electron)} & \textbf{$\sim$19\,\%} & \textbf{$\sim$67\,\%} \\
\hline
\end{tabularx}
\end{center}
\end{table}

\subsection{Polarization Dependence}

Most waveguides are birefringent because the waveguide cross-section is not
perfectly symmetric, so the transverse-electric and transverse-magnetic modes
have different effective indices. At $R\!>\!20{,}000$ this birefringence
separates the two polarization channels enough to broaden the composite line
and degrade resolution for unpolarized light. The two remedies are to split
the input polarization and feed two copies of a polarization-optimised AWG,
or to engineer the birefringence out of the waveguide through cross-section
design, width variation along the array, or a slanted interface between the
free-propagation region and the waveguide array. Polarization control is also
the enabling requirement for the spectropolarimetry driver of
Section~\ref{sect:pic-applications}.

\subsection{Extending AWGs to the Ultraviolet}

Silica and silicon-nitride AWGs serve the optical and near-infrared. UV AWGs
require the wide-bandgap platforms of Section~\ref{sect:materials}, such as
aluminium nitride, aluminium oxide, and tantalum pentoxide, together with
low-roughness patterning to control Rayleigh scattering. UV AWGs have been
demonstrated, but their throughput, resolving power, and bandwidth must all
be advanced substantially before they can serve the HWO Growth of Galaxies and
Evolution of the Elements drivers.

It is important to state the practical short-wavelength limits honestly.
For the Al$_2$O$_3$ platform, recent progress is notable: atomic-layer-deposited
(ALD) alumina waveguides have demonstrated sub-dB/cm propagation losses at
$360\,\mathrm{nm}$, and a fully CMOS-fabricated piezo-optomechanical platform
using ALD alumina cores on $200\,\mathrm{mm}$ wafers has demonstrated a
reconfigurable racetrack-resonator filter operating at $320\,\mathrm{nm}$
(propagation loss $\leq\!4.4\,\mathrm{dB/cm}$, $6\,\mathrm{ns}$ switching
time), with the platform claimed to be compatible with operation down to
$225\,\mathrm{nm}$~\citep{Castillo2026AluminaUV}.
Nitride platforms, in particular AlN and AlGaN on sapphire, are the other
candidate route into the ultraviolet, with waveguides, microring resonators
and filters reported across the near-ultraviolet~\citep{Blumenthal2020}.
Both platforms thus establish a credible path toward HWO's near-UV requirements,
but a critical distinction must be noted: no AWG structure has yet been
demonstrated in the Al$_2$O$_3$ CMOS platform, and none in any UV platform has
yet reached the combination of resolving power ($R\!\gtrsim\!10^{4}$), insertion
loss ($<\!3\,\mathrm{dB}$), and bandwidth ($>\!20\,\mathrm{nm}$) that HWO
science requires.
At present, the most capable UV PIC demonstrations are ring and filter
structures in Al$_2$O$_3$, Ta$_2$O$_5$ and the nitrides, with the shortest
published operating wavelength for a fully integrated, CMOS-fabricated device
being the $320\,\mathrm{nm}$ alumina racetrack filter cited
above~\citep{Blumenthal2020,Castillo2026AluminaUV}.
The far-ultraviolet range below $\sim$200\,nm, which encompasses the
hydrogen Lyman-$\alpha$ line and several biosignature bands important to the
Living Worlds and Growth of Galaxies working groups (Table~\ref{tab:drivers}), is
not yet reachable with any demonstrated PIC AWG platform; guided-wave
operation at these wavelengths requires bulk-material transparency, waveguide
sidewall quality, and fibre-coupling solutions that remain at the research
stage~\citep{Blumenthal2020}.
This gap is the most critical UV materials challenge for HWO astrophotonics:
it should be addressed explicitly in Phase~1 of the roadmap (Section~\ref{sect:roadmap}),
with the target of demonstrating an AlN AWG at $\lambda\!\lesssim\!250\,\mathrm{nm}$
with $R\!\geq\!1000$ and $<\!3\,\mathrm{dB}$ insertion loss.

\subsection{The Potsdam Arrayed Waveguide Spectrograph (PAWS)}
\label{subsect:paws}

The Potsdam Arrayed Waveguide Spectrograph (PAWS) is a laboratory demonstrator that established end-to-end operation of a fibre-coupled silica-on-silicon AWG spectrograph~\citep{Stoll2020PAWS,Stoll2017PAWS}.
PAWS accepts light from a single-mode optical fibre, disperses it across five spectral orders in the near-infrared $H$-band (1.48--1.74\,$\mu$m), and records the resulting echellogram on a Teledyne Hawaii\,2RG detector array, reaching resolving powers up to $R\!\sim\!10{,}000$~\citep{Stoll2020PAWS}.
The instrument demonstrates the complete signal chain of a photonic spectrograph, fibre coupling, guided-wave dispersion, order separation by a bulk cross-disperser, and detector read-out, and confirms that single-mode AWG devices achieve the throughput and resolving power predicted by design.

\begin{figure}
\centering
\includegraphics[width=0.72\linewidth]{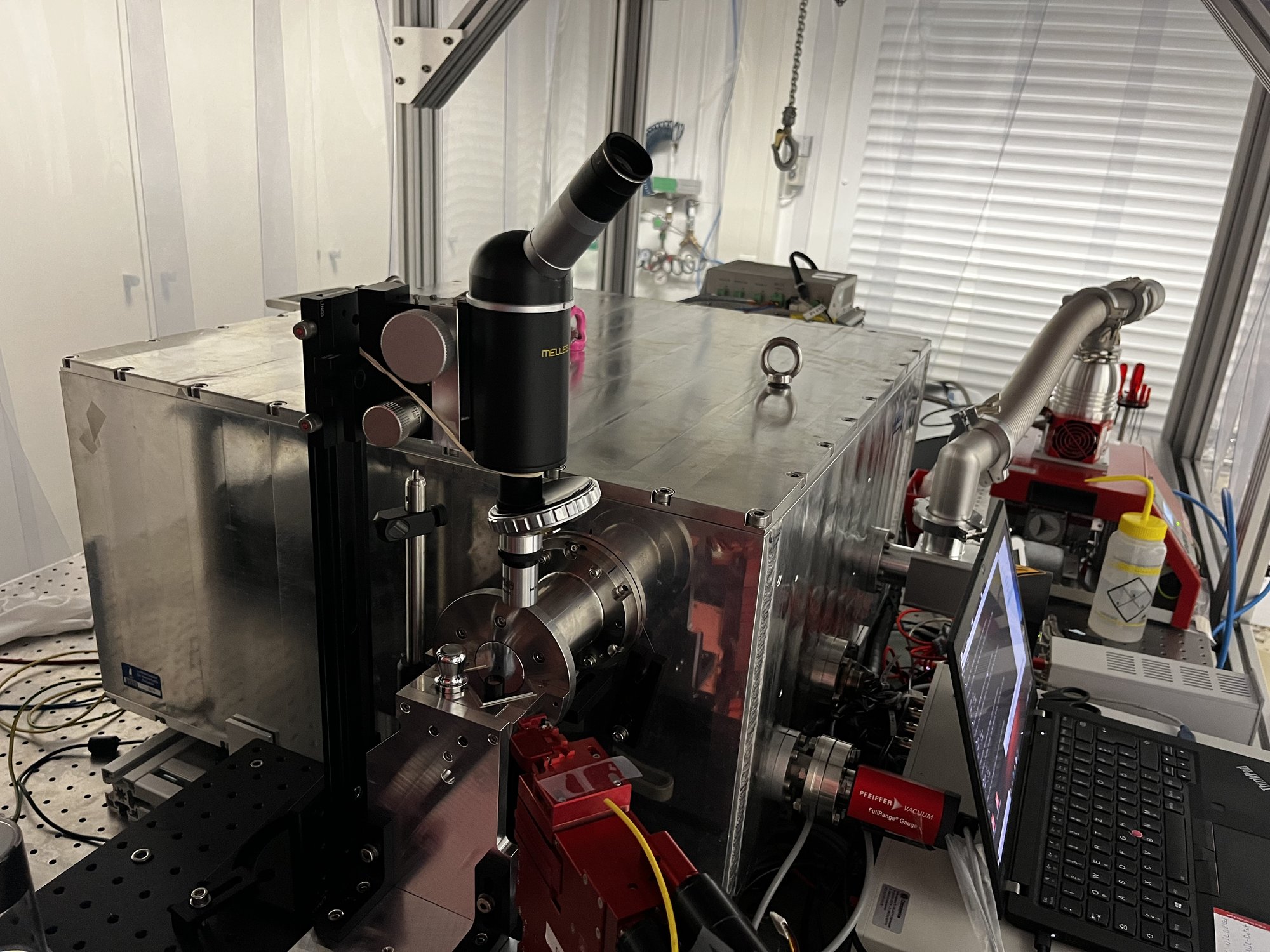}
\caption{The Potsdam Arrayed Waveguide Spectrograph (PAWS) laboratory demonstrator~\citep{Stoll2020PAWS}.
The silica-on-silicon AWG chip and its single-mode fibre feed are mounted \emph{outside} the cryostat;
an infinite objective focuses the AWG output through a sapphire vacuum window into the cryostat interior.
Inside, the light is cross-dispersed by a bulk reflection grating to separate the five spectral orders
(H-band, $\lambda = 1.48$--$1.74\,\mu$m, resolving power $R\!\sim\!10{,}000$) and recorded on a
Teledyne Hawaii\,2RG near-infrared detector array cooled to $\sim$77\,K to suppress dark current.
The cross-disperser mount and detector assembly on the cold plate are visible in Fig.~\ref{fig:cold_plate_paws}.
PAWS is a ground-based laboratory demonstrator; it is not designed or intended as a space instrument,
but it validates the complete photonic spectrograph signal chain described in Section~\ref{subsect:paws}.}
\label{fig:paws_demonstrator}
\end{figure}

\begin{figure}
\centering
\includegraphics[width=0.90\linewidth]{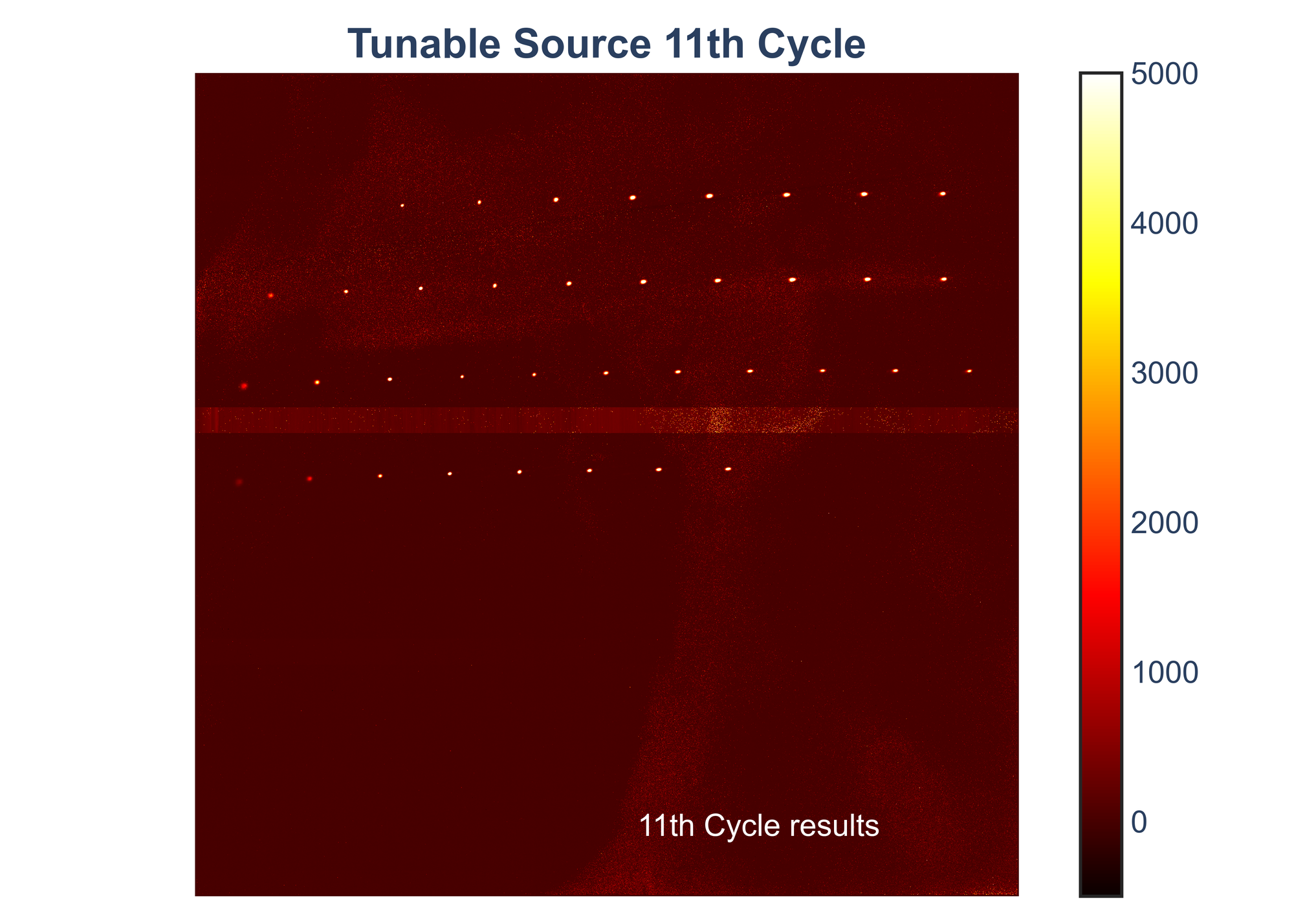}
\caption{Near-infrared echellogram recorded by the PAWS demonstrator (Teledyne Hawaii\,2RG detector) when the AWG is illuminated with a tuneable laser source in the wavelength range $1480$--$1741\,$nm.  The five spectral orders of the Type-F AWG are separated vertically by the bulk cross-disperser grating and dispersed horizontally by the AWG, producing the characteristic ladder of spectral stripes.}
\label{fig:paws_frame}
\end{figure}

The spectral output of the PAWS instrument is shown in Fig.~\ref{fig:paws_frame}, which displays the near-infrared echellogram recorded on the Hawaii\,2RG detector when the AWG is illuminated with a tuneable laser source in 5\,nm wavelength steps.

\begin{figure}
\centering
\includegraphics[angle=-90, width=0.65\linewidth]{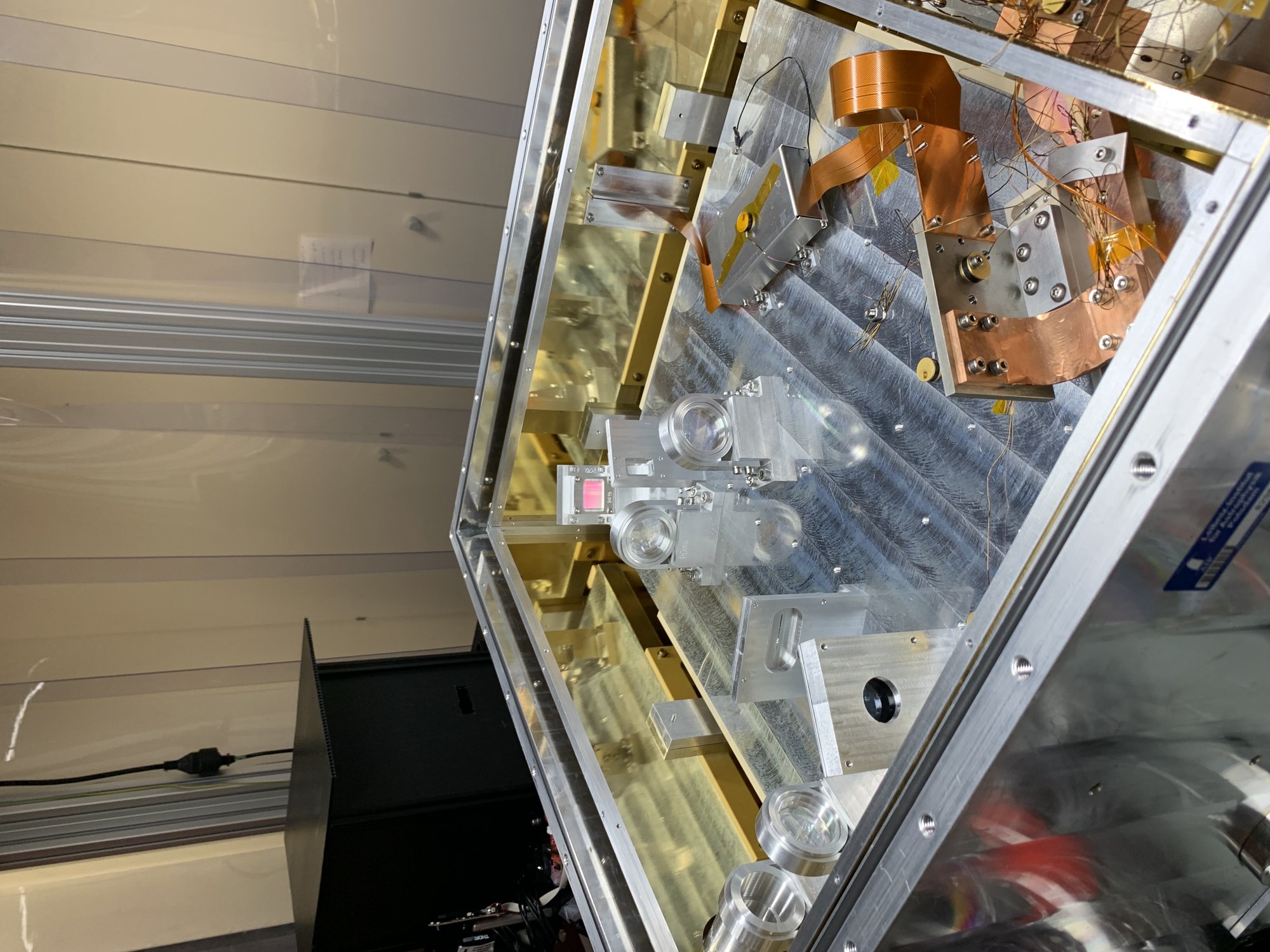}
\caption{Cryogenic cold plate of the PAWS laboratory demonstrator at $\sim$77\,K, showing the cold stops, cross-disperser, relay optics, and Teledyne Hawaii\,2RG detector interface within the cryostat cold stage. The AWG chip and fibre feed are located outside the cryostat; light enters through a sapphire vacuum window (not visible here) via an infinite objective.}
\label{fig:cold_plate_paws}
\end{figure}

The cryogenic mounting architecture of PAWS is shown in Fig.~\ref{fig:cold_plate_paws}. The cold plate is cooled to approximately $77\,\mathrm{K}$ to suppress detector dark current in the Hawaii\,2RG array; the silica-on-silicon AWG chip and fibre feed are mounted outside in the prototype with an infinite objective passing the light into the cryostat through a sapphire window. This arrangement was deliberate: it allows different AWG architectures and fibre feeds to be tested without a full cryostat turnaround, including warm-up, venting, pump-down, and re-cooling.

\begin{figure}
\centering
\includegraphics[width=0.95\linewidth]{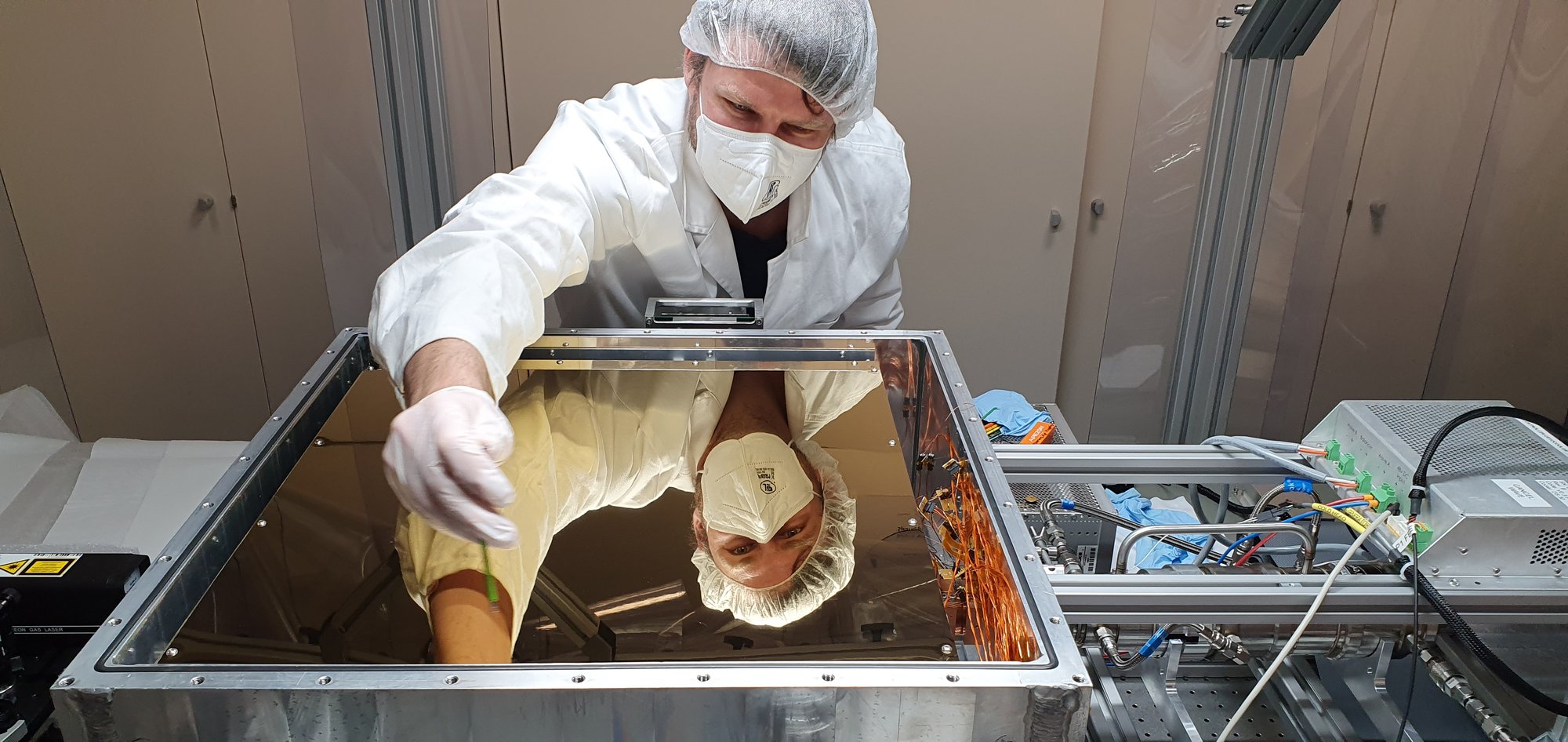}
\caption{PAWS laboratory demonstrator enclosure with the top cover removed, showing the top radiation shield.}
\label{fig:paws_top_open}
\end{figure}

The internal layout of the PAWS cryostat is shown in Fig.~\ref{fig:paws_top_open} with the top radiation shield removed, revealing the fibre-ribbon entry feedthrough, the cross-disperser stage, and the detector assembly.

\subsection{Size and Mass Advantage of PIC Spectrographs}
\label{subsect:pic_mass}

A key motivator for PIC spectrographs on space missions is the dramatic reduction in size, mass, and mechanical complexity relative to conventional bulk-optic instruments.
Table~\ref{tab:mass_comparison} quantifies this advantage by comparing two representative state-of-the-art space near-infrared spectrographs against a projected PIC-based NIR spectrograph module built from stacked silica-on-silicon AWGs with fibre-ribbon inputs and embedded Teledyne space-qualified detectors.

JWST NIRSpec covers 0.6--5.3\,$\mu$m and provides multi-object spectroscopy for up to $\sim$100 simultaneous targets via its microshutter assembly~\citep{Jakobsen2022NIRSpec}.
The complete instrument has a mass of approximately 196\,kg and occupies an envelope of $\sim$$1.9\!\times\!0.9\!\times\!0.7\,\mathrm{m}$~\citep{Jakobsen2022NIRSpec}.
MAJIS (Moons And Jupiter Imaging Spectrometer) aboard ESA's JUICE mission covers 0.5--5.7\,$\mu$m with two spectral channels and represents a compact, single-slit imaging spectrometer; the instrument mass is approximately 23\,kg~\citep{Poulet2024MAJIS}.
Both instruments use conventional bulk-optic architectures: dispersing gratings or prisms, relay optics, and separate detector assemblies mounted on thermally stabilised benches.

For a PIC-based NIR spectrograph, the chip-scale parameter $X \equiv R \times \mathrm{FSR} / \lambda_0$ provides a figure of merit for the per-order information content of a single AWG channel~\citep{Stoll21}.
For silica-on-silicon AWGs, $X \approx 370$, and at a centre wavelength of $\lambda_0 = 1550\,\mathrm{nm}$ this yields a single-order design with $R \approx 1{,}900$ and $\mathrm{FSR} \approx 300\,\mathrm{nm}$, covering a 300\,nm passband in a single spectral order with no cross-dispersion required.
A silica AWG chip implementing this design has a footprint of approximately $30\!\times\!30\,\mathrm{mm}$ and a substrate thickness of $\sim$1\,mm~\citep{Stoll2020PAWS,Stoll21}.
To cover the NIR range 0.95--2.75\,$\mu$m (J$+$H$+$K bands) five such chips are required, each shifted in centre wavelength by 300\,nm.
The five chips may be stacked or tiled in a $30\!\times\!30\!\times\!5\,\mathrm{mm}$ assembly with a mass of order 10\,g for the AWG substrates alone.
At each chip's output facet, a space-qualified detector array reads the dispersed spectrum directly.
Two complementary Teledyne detector families are relevant, covering different wavelength ranges and pixel-pitch requirements.

For near-infrared wavelengths ($\gtrsim\!0.9\,\mu$m), Teledyne's H2RG HgCdTe arrays ($2048\!\times\!2048$ pixels, $18\,\mu$m pitch, mass $\sim\!30\,\mathrm{g}$ per unit) are flight-qualified for JWST and provide the broad-band NIR sensitivity required for J$+$H$+$K coverage~\citep{Blank2012H2RG}.

For the visible and near-UV range, and for higher-density pixel coupling to fine-pitch SiN AWG output channels, Teledyne's space-qualified CMOS Time-Delay Integration (TDI) sensors offer a complementary capability.
The IC-52-12K2 is a representative example: a hybrid charge-domain TDI/CMOS device that combines CCD-grade charge transfer efficiency (CTE $\geq\!0.9999$ per stage) with CMOS readout and on-chip 12-bit analogue-to-digital conversion on a single monolithic chip, available in both front-side-illuminated (FSI) and back-side-illuminated (BSI) variants~\citep{TeledyneIC52}.
The BSI variant achieves $4$--$5\times$ higher responsivity than the FSI version across the six multispectral bands (B1--B6).
Key specifications relevant to AWG spectroscopy are summarised in Table~\ref{tab:teledyne_cmos}: a $7\,\mu\mathrm{m}\!\times\!7\,\mu\mathrm{m}$ pixel pitch, 12\,288 pixels per spectral band, full-well capacity $\geq\!85{,}000\,\mathrm{e}^{-}$, read noise $\leq\!60\,\mathrm{e}^{-}$ RMS (FSI), and a total power dissipation $\leq\!10\,\mathrm{W}$~\citep{TeledyneIC52}.
Space qualification is explicitly addressed in the IC-52-12K2 datasheet: total ionising dose (TID) tolerance $\geq\!20\,\mathrm{krad(Si)}$ (Co-60) and no destructive single-event latch-up (SEL) at linear energy transfers $\geq\!60\,\mathrm{MeV\,mg^{-1}\,cm^{2}}$, placing it within the planning range for the Sun--Earth $L_2$ radiation environment discussed in Section~\ref{sect:space}~\citep{TeledyneIC52}.

The $7\,\mu$m pixel pitch is directly compatible with the output channel spacings achievable in high-index-contrast SiN AWG platforms fabricated with deep-UV lithography, potentially enabling butt-coupling without relay optics for SiN devices.
For the silica-on-silicon platform, where output channel spacings are typically $20$--$100\,\mu$m, simple demagnifying relay optics ($3$--$15\times$) bridge the pitch mismatch at negligible volume cost relative to the AWG chip itself.

\begin{table}
\caption{Key specifications of the Teledyne IC-52-12K2 CMOS TDI sensor relevant to AWG spectrograph coupling~\citep{TeledyneIC52}. FSI = front-side illuminated; BSI = back-side illuminated. TID = total ionising dose; SEL = single-event latch-up. Where FSI and BSI values differ, both are given in a single cell as ``FSI value; BSI value''.}
\label{tab:teledyne_cmos}
\begin{center}
\footnotesize
\begin{tabularx}{\linewidth}{|X|c|}
\hline
\rule[-1ex]{0pt}{3.5ex}
\textbf{Parameter} & \textbf{IC-52-12K2 (FSI / BSI)} \\
\hline\hline
\rule[-1ex]{0pt}{3.5ex}
Pixel size & $7\,\mu\mathrm{m} \times 7\,\mu\mathrm{m}$ \\
\hline
\rule[-1ex]{0pt}{3.5ex}
Pixels per spectral band & 12\,288 \\
\hline
\rule[-1ex]{0pt}{3.5ex}
Spectral bands & 6 (B1--B6, multispectral; BSI version) \\
\hline
\rule[-1ex]{0pt}{3.5ex}
CTE per stage & $\geq\!0.9999$ \\
\hline
\rule[-1ex]{0pt}{3.5ex}
Full well capacity & $\geq\!85{,}000\,\mathrm{e}^{-}$ \\
\hline
\rule[-1ex]{0pt}{3.5ex}
Read noise RMS & $\leq\!60\,\mathrm{e}^{-}$ (FSI) \\
\hline
\rule[-1ex]{0pt}{3.5ex}
On-chip ADC & 12-bit \\
\hline
\rule[-1ex]{0pt}{3.5ex}
Dark current density (25\,$^\circ$C) & $\leq\!6\,\mathrm{nA\,cm^{-2}}$ (FSI); $\leq\!12\,\mathrm{nA\,cm^{-2}}$ (BSI) \\
\hline
\rule[-1ex]{0pt}{3.5ex}
Power dissipation & $\leq\!10\,\mathrm{W}$ \\
\hline
\rule[-1ex]{0pt}{3.5ex}
Radiation tolerance (TID) & $\geq\!20\,\mathrm{krad(Si)}$, Co-60 \\
\hline
\rule[-1ex]{0pt}{3.5ex}
SEL immunity & No latch-up at LET $\geq\!60\,\mathrm{MeV\,mg^{-1}\,cm^{2}}$ \\
\hline
\rule[-1ex]{0pt}{3.5ex}
Pixel pitch match to SiN AWGs & Direct butt-coupling possible \\
\hline
\end{tabularx}
\end{center}
\end{table}

Including fibre-ribbon coupling, detector assemblies, and a compact housing, a single-target PIC NIR spectrograph module covering 0.95--2.75\,$\mu$m is projected to have a total mass of order $0.3$--$0.5\,\mathrm{kg}$ and a volume below $100\,\mathrm{cm}^3$.
Because each module handles one fibre input, multi-object capability scales by replication: ten modules provide 10-target MOS at $\sim$3--5\,kg total; 100 modules provide NIRSpec-class multiplexing at $\sim$30--50\,kg, in a fraction of the NIRSpec volume and with minimal moving parts.

\begin{table}
\caption{Indicative size and mass comparison of conventional space NIR spectrographs and a \emph{projected} PIC silica-on-silicon based AWG spectrograph module. PIC module values are engineering estimates based on current silica-on-silicon chip dimensions, published AWG chip footprints~\citep{Stoll2020PAWS,Stoll21}, and space-qualified detector masses~\citep{Blank2012H2RG}; they have not been demonstrated in hardware. A space-qualified housing would add $\sim$0.2\,kg per module.}
\label{tab:mass_comparison}
\begin{center}
\footnotesize
\begin{tabularx}{\linewidth}{|X|c|c|X|X|}
\hline
\rule[-1ex]{0pt}{3.5ex}
\textbf{Parameter} & \textbf{NIRSpec} & \textbf{MAJIS} & \textbf{PIC module} & \textbf{PIC $\times$100} \\
& \textbf{(JWST)} & \textbf{(JUICE)} & \textbf{(\emph{projected})} & \textbf{(\emph{projected})} \\
\hline\hline
\rule[-1ex]{0pt}{3.5ex}
Wavelength range & 0.6--5.3\,$\mu$m & 0.5--5.7\,$\mu$m & 0.95--2.75\,$\mu$m & 0.95--2.75\,$\mu$m \\
\hline
\rule[-1ex]{0pt}{3.5ex}
Resolving power & 100--2700$^b$ & 1000--3000 & $\sim$1900 & $\sim$1900 \\
\hline
\rule[-1ex]{0pt}{3.5ex}
Simultaneous targets & $\sim$100 (MOS) & 1 (slit) & 1 & 100 \\
\hline
\rule[-1ex]{0pt}{3.5ex}
Instrument mass & $\sim$196\,kg & $\sim$23\,kg & $\sim$0.3--0.5\,kg & $\sim$30--50\,kg \\
\hline
\rule[-1ex]{0pt}{3.5ex}
Volume envelope & $1.9\!\times\!0.9\!\times\!0.7$\,m & $0.4\!\times\!0.3\!\times\!0.3$\,m & $<100\,\mathrm{cm}^{3}$ & $<0.01\,\mathrm{m}^{3}$ \\
\hline
\rule[-1ex]{0pt}{3.5ex}
Moving parts & grating wheel & filter wheel & none & none \\
\hline
\rule[-1ex]{0pt}{3.5ex}
Cross-dispersion & bulk grating & prism & required for $R$>1900 $^a$ & required for $R$>1900 $^a$ \\
\hline
\end{tabularx}
{\raggedright $^a$ The size of the cross-disperser grating and imaging optics is proportional to the size of the PIC, and does not scale with the number of PICs in MOS designs.\\
$^b$ NIRSpec $R\!\approx\!100$ (prism), 1000, and 2700 (grating modes)~\citep{Jakobsen2022NIRSpec}; AWG targets the complementary high-$R$ ($\sim\!10^{4}$--$10^{5}$) regime. \par}
\end{center}
\end{table}

The mass advantage is not merely a size-reduction benefit: the elimination of the grating wheel mechanism (Section~\ref{sect:pic-applications}), collimating and camera optics, and the thermally stabilised optical bench removes the dominant mass, volume, and failure-mode drivers of conventional spectrographs.
For HWO, where each kilogram of instrument mass carries a launch cost and each mechanism represents a reliability risk over a decade-plus mission lifetime, the PIC architecture offers a substantial and quantifiable advantage in mass, volume, and mechanism count at the $10^{3}$--$10^{4}$ channel scale, provided that the space-qualification requirements identified in Section~\ref{sect:space} are met.

\subsection{Detector Integration at the AWG Output Facet}
\label{subsect:detector_int}

The projected size and mass advantage of Section~\ref{subsect:pic_mass} is fully realised only if the detector assembly does not reintroduce the bulk-optic relay and alignment interfaces that the AWG chip eliminates.
The detector interface is therefore a first-order design problem for any AWG spectrograph.
Three issues dominate: (i) spatial pitch matching between the AWG output channel spacing and the detector pixel pitch; (ii) coefficient-of-thermal-expansion (CTE) mismatch between the AWG substrate, any relay optics, and the detector material at the cryogenic operating temperature; and (iii) the quantum efficiency (QE), read noise, and dark current of the detector, which together determine how much of the optical efficiency in Table~\ref{tab:throughput} is ultimately converted to signal.
In the conventional approach, AWG output light exits the chip edge and is collected by relay optics onto a separate detector array (e.g.\ Teledyne H2RG), incurring the 80\,\% relay-optics coupling loss listed in Table~\ref{tab:throughput} and introducing a mechanically sensitive alignment interface that must survive cryogenic cycling and launch vibration.
Attaching the detector directly at the output facet, by butt-coupling or flip-chip bonding, removes this interface.

Direct attachment is favoured by the way an AWG presents its output. The channel spacing is fixed lithographically rather than by an assembly tolerance, so a detector whose pixel pitch matches that spacing can be coupled without demagnifying relay optics; the $7\,\mu$m pitch of the CMOS TDI sensor of Section~\ref{subsect:pic_mass} is already within reach of the output spacings achievable on high-index-contrast silicon-nitride platforms. Silicon nitride itself provides a low-loss, wide-bandgap optical bus from $\sim$0.3\,$\mu$m to $\sim$2.5\,$\mu$m, covering the full near-infrared range of HWO science drivers 1--3 (Table~\ref{tab:drivers}), so the wavelength reach of a directly coupled spectrograph is set by the detector material rather than by the waveguide. Longer-wavelength coverage is addressed by bonding HgCdTe or extended-InGaAs detectors at the chip facet.

What direct attachment does not remove is the thermal and noise problem, and the requirements there are set by astronomy rather than by the telecommunications applications that have driven detector integration to date. Photon rates are low and integration times long, so dark current and read noise dominate rather than bandwidth. The dark-current values in Table~\ref{tab:teledyne_cmos} are room-temperature (25\,$^\circ$C) datasheet figures; the corresponding cryogenic figures, which are the ones that matter for HWO, must be measured for any candidate detector at its operating temperature, together with the evolution of dark current and read noise under the radiation dose expected at the Sun--Earth $L_2$ point. Coefficient-of-thermal-expansion mismatch likewise becomes a bonding problem to be qualified across the $70$--$300\,\mathrm{K}$ range rather than an alignment problem to be maintained, which is how it appears in the requirement-to-test mapping of Section~\ref{sect:trl}.

Direct detector attachment therefore addresses two of the three interface issues, pitch matching and relay-optics loss, and converts the third into a qualification requirement. When combined with the single-order AWG design of Section~\ref{subsect:pic_mass} (no cross-dispersion required), it removes the two largest loss terms in Table~\ref{tab:throughput} and projects an end-to-end photon-to-electron efficiency approaching 67\,\% in the optimised case, a factor of $\sim$3.6 improvement over the current unoptimised baseline and equivalent, in the photon budget, to a factor of 3.6 in collecting area, which no aperture choice within the EAC range can supply.

\section{Space Qualification of PICs and Optical Fibres}
\label{sect:space}

\subsection{The Space Environment}

A photonic instrument bound for the Sun--Earth $L_2$ point must survive four
classes of environmental stress: ionising radiation (trapped particles,
cosmic rays, and solar events); thermal cycling and operation at low or
cryogenic temperature; hard vacuum; and the vibration and shock of launch.
None of these is fully addressed by the laboratory and on-sky demonstrations
that have established the optical performance of PICs and fibres to date.
Space qualification is therefore the principal programmatic gap between the
current state of the art and an HWO photonic instrument. The present section
sets out the environmental drivers and the resulting requirements at the level
needed to support the TRL assessment of Section~\ref{sect:trl}; a detailed
test-by-test qualification template for astrophotonic components, consolidating
the applicable NASA and ESA guidelines, is developed in a companion
paper~\citep{Madhav2026PICSQT}.

\subsection{Radiation Effects}

Ionising radiation creates point defects or colour centres in glasses,
producing radiation-induced absorption that darkens fibres and waveguides.
The effect is strongly wavelength-dependent and most severe in the
ultraviolet, where it compounds the solarization that also affects UV-grade
silica~\citep{Girard2019}. Radiation hardness depends strongly on material
composition, so each candidate waveguide and fibre platform must be
characterised individually. Hollow-core fibres, which guide light in air,
are intrinsically less susceptible to radiation-induced absorption and are an
attractive mitigation for UV light paths. Quantitative, material-by-material
radiation testing against the expected $L_2$ dose is a required development
step.

\subsection{Thermal Cycling and Cryogenic Operation}

Near-infrared astronomical detectors operate near $70\,\mathrm{K}$, and a
PIC butt-coupled or bonded to such a detector must tolerate the same
temperature. The dominant risk is the mismatch in coefficient of thermal
expansion (CTE) between dissimilar materials of waveguide, substrate,
fibre, adhesive, and detector, which induces stress and can crack the
fibre-to-chip bond. The pigtailed germanium-doped silica beam combiners of
GRAVITY have been operated successfully at $-80\,^{\circ}\mathrm{C}$, which
establishes partial heritage~\citep{GRAVITY2017}, but HWO will require
qualification of bonded PIC--fibre--detector assemblies across the full
thermal range and many cycles.
The key interface engineering challenge is the mounting of the AWG chip, fibre ribbon, and detector array on a common thermally conductive structure that minimises differential CTE stress while providing a controlled thermal path to the cryocooler cold finger.

\subsection{Vacuum, Outgassing, and Bonding}
\label{subsect:vacuum}

In hard vacuum, index-matching gels and oils are unsuitable: they outgas and deposit on cold optical surfaces, and their mechanical properties change unpredictably through the cryogenic temperature range.
A space-qualified PIC instrument therefore requires solid, durable, crack-free
fibre-to-chip and chip-to-chip bonds achieved without index-matching fluids.
Developing such bonds, and qualifying them against thermal cycling and
vibration, is a central packaging challenge shared with the broader effort to
build hybrid photonic instruments.

\subsection{Heritage and a Qualification Pathway}

Photonic components have partial spaceflight and high-altitude heritage:
UV-transmitting fibres flew on the FIREBall balloon experiment, and photonic
beam combiners operate routinely under the cryogenic, vibration-prone
conditions of ground-based interferometry. A credible HWO qualification
pathway builds on this heritage in three steps.
First, component-level
environmental testing (radiation, thermal-vacuum, and vibration) for each
waveguide platform, fibre type, and bond. Second, subsystem-level testing of
a complete photonic spectrograph or calibration unit. Third, a technology-
demonstrator flight on a balloon, sounding rocket, or small satellite, which
both retires risk and builds the flight heritage that mission reviews
require. The individual test steps, acceptance criteria and TRL gates that
populate these three stages are specified in the companion qualification
template~\citep{Madhav2026PICSQT}. This pathway should be pursued in parallel
with the optical-performance development of Sections~\ref{sect:materials} and
\ref{sect:awg}, not after it.

\section{Key Technical Requirements and Technology Readiness Levels}
\label{sect:trl}

Table~\ref{tab:trl} consolidates the key technical requirements derived in
Sections~\ref{sect:pic-applications}--\ref{sect:space}, with an assessment of
the current Technology Readiness Level (TRL) and the TRL required at HWO's
instrument-definition gate. TRL values follow the NASA/ESA nine-level scale~\citep{Mankins1995TRL}.
The assessments are based on the literature cited in the corresponding sections
(on-sky demonstrations and laboratory publications for optical/NIR platforms,
primarily laboratory-stage results for UV platforms and space-qualification
categories), and are necessarily approximate and platform-dependent; they are
intended to identify the largest development gaps rather than to provide a
formal programmatic audit.

\begin{table*}
\caption{Key technical requirements for HWO photonic instruments, with
current and required Technology Readiness Levels (TRLs). TRL values are
approximate and platform-dependent.}
\label{tab:trl}
\begin{center}
\footnotesize
\begin{tabularx}{\textwidth}{|l|X|c|c|}
\hline
\rule[-1ex]{0pt}{3.5ex}
\textbf{Area} & \textbf{Key technical requirement} & \textbf{Current TRL} &
\textbf{HWO need} \\
\hline\hline
\multicolumn{4}{|l|}{\textit{Material platforms}} \\
\hline
\rule[-1ex]{0pt}{3.5ex}
UV waveguides & AlN / Al$_2$O$_3$ / Ta$_2$O$_5$ waveguides with
$<\!1\,\mathrm{dB\,cm^{-1}}$ loss from $\sim\!100$--$400\,\mathrm{nm}$ &
2--3 & 5--6 \\
\hline
\rule[-1ex]{0pt}{3.5ex}
Optical/NIR waveguides & Si$_3$N$_4$ / silica with
$<\!0.1\,\mathrm{dB\,cm^{-1}}$ loss over $0.4$--$2.5\,$\textmu m &
4--5 & 6 \\
\hline
\rule[-1ex]{0pt}{3.5ex}
Optical fibres & UV-grade and solarization-resistant hollow-core fibres
spanning the HWO band & 3--4 & 6 \\
\hline
\multicolumn{4}{|l|}{\textit{AWG spectrographs}} \\
\hline
\rule[-1ex]{0pt}{3.5ex}
High-$R$ AWGs & On-chip $R\!\geq\!10^{5}$ via tandem cross-dispersion;
path-length control $\sim\!10\,\mathrm{ppm}$ & 4 & 6 \\
\hline
\rule[-1ex]{0pt}{3.5ex}
Throughput & End-to-end throughput $\gtrsim\!50\,\%$ via optimised tapers,
AR coatings, cascaded AWGs & 3--4 & 6 \\
\hline
\rule[-1ex]{0pt}{3.5ex}
Polarization & Birefringence-compensated or polarization-diverse AWGs for
unpolarized and spectropolarimetric use & 3--4 & 5--6 \\
\hline
\rule[-1ex]{0pt}{3.5ex}
UV AWGs & UV-platform AWGs with astronomical throughput, bandwidth, and
resolving power & 2--3 & 5 \\
\hline
\multicolumn{4}{|l|}{\textit{Calibration and active components}} \\
\hline
\rule[-1ex]{0pt}{3.5ex}
On-chip frequency combs & Microresonator combs, $\geq\!300\,\mathrm{nm}$
bandwidth, multi-year stability & 4 & 6 \\
\hline
\rule[-1ex]{0pt}{3.5ex}
Fabry--P\'erot references & On-chip etalons as stable secondary wavelength
references & 4--5 & 6 \\
\hline
\multicolumn{4}{|l|}{\textit{Integration and space qualification}} \\
\hline
\rule[-1ex]{0pt}{3.5ex}
Fibre-to-chip coupling & $>\!95\,\%$ ($<\!0.4\,\mathrm{dB}$) broadband
coupling without index-matching fluids & 3--4 & 6 \\
\hline
\rule[-1ex]{0pt}{3.5ex}
Detector integration & Butt-coupled or flip-chip detector bonding to AWG
focal plane & 3--4 & 5--6 \\
\hline
\rule[-1ex]{0pt}{3.5ex}
Radiation tolerance & Material-by-material characterisation against the
$L_2$ radiation dose & 2--3 & 6 \\
\hline
\rule[-1ex]{0pt}{3.5ex}
Thermal / vacuum / vibration & Qualified PIC--fibre--detector bonds across
$70$--$300\,\mathrm{K}$, vacuum, and launch loads & 3 & 6 \\
\hline
\end{tabularx}
\end{center}
\end{table*}

Three conclusions follow from Table~\ref{tab:trl}. First, the optical/near-
infrared waveguide and AWG technologies are the most mature (TRL 4--5):
their gap to HWO is one of optimisation and qualification rather than
invention. Second, ultraviolet waveguides and AWGs are the least mature
(TRL 2--3) and require dedicated materials development. Third, space
qualification (radiation, thermal, vacuum, and vibration) sits at
TRL 2--3 across the board and is the most uniformly under-addressed area.
It is also the longest-lead activity and should therefore start earliest.

\section{A Development Roadmap for HWO Photonic Instruments}
\label{sect:roadmap}

A three-phase development programme is proposed, spanning the approximately
fifteen years between now and HWO's instrument-definition gate (notionally
$\sim$2038--2040 based on current NASA planning timelines; all dates are
indicative and subject to mission schedule revision). The phases
overlap; in particular, space-qualification activities (Phase~3 emphasis)
must begin during Phase~1 because they are the longest-lead items.
No specific resource or staffing levels are assumed here; the phases describe
technical milestones rather than a costed programme plan.

\noindent\textit{Phase 1 ($\sim$2026--2030): materials and components.}
Advance UV waveguide platforms (AlN, Al$_2$O$_3$, Ta$_2$O$_5$) to TRL\,4
with characterised loss budgets, including a first AlN AWG demonstration at
$\lambda\!\lesssim\!250\,\mathrm{nm}$ (Section~\ref{sect:awg});
deliver an optical/near-infrared tandem
AWG spectrograph at $R\!\geq\!10^{4}$ on a ground-based telescope; mature
on-chip frequency combs to $\geq\!300\,\mathrm{nm}$ bandwidth; and begin
component-level radiation and thermal-vacuum testing of each candidate
waveguide and fibre material.
In parallel, the dark current, read noise and cryogenic performance of candidate detectors for direct attachment at an AWG output facet (Section~\ref{subsect:detector_int}) should be characterised at their intended operating temperature, and a directly coupled AWG spectrograph demonstrated at laboratory level.

\noindent\textit{Phase 2 ($\sim$2030--2035): subsystems and qualification.}
Demonstrate a fully integrated photonic spectrograph with fibre feed,
SiN AWG, and directly coupled detector at $R\!\sim\!10^{4}$--$10^{5}$ with
end-to-end photon-to-electron efficiency above fifty percent (Section~\ref{subsect:detector_int}); demonstrate a UV PIC
spectrograph at $R\!\sim\!10^{4}$; qualify fibre-to-chip and chip-to-chip
bonds across the $70$--$300\,\mathrm{K}$ range and against vibration; characterise radiation tolerance of SiN waveguides and the coupled detector against the expected $L_2$ dose; and
fly a photonic spectrograph or calibration unit on a balloon or sounding
rocket ($\sim$2032--2034 target) to build flight heritage.

\noindent\textit{Phase 3 ($\sim$2035--2040): mission integration.}
Deliver TRL-6 photonic spectrograph, integral-field, calibration, and
coronagraph-back-end subsystems for incorporation into HWO instrument
proposals; complete full environmental qualification of the integrated
PIC--fibre--detector assemblies; and establish multi-project-wafer
fabrication runs so that the multi-object-spectroscopy channels can be
replicated at the required scale.

Success depends on close collaboration between academic groups (well placed
to advance designs and to qualify devices), the integrated-photonics industry
(well placed to provide fabrication and packaging at scale), and the space
agencies (which define the mission context and qualification standards).
Multidisciplinary centres that combine astronomical instrument expertise with
photonic fabrication, such as Astrophotonics (innoFSPEC) at AIP Potsdam, are
positioned to coordinate such a programme.

\section{Conclusions}
\label{sect:conclusions}

Photonic integrated circuits and optical fibres are examined as enabling
technologies for the Habitable Worlds Observatory. From the HWO25 Science
Case Development Documents six observing-mode drivers are identified: 
coronagraph-fed spectroscopy, high-resolution cross-correlation spectroscopy,
multi-object UV spectroscopy, integral-field spectroscopy, extreme-precision
radial velocity, and UV spectropolarimetry, for which photonic
implementations are advantageous or enabling. Each driver is mapped onto a
concrete PIC application, showing that covering the
$100\,\mathrm{nm}$-to-$2.5\,$\textmu m HWO range requires a set of
complementary waveguide and fibre materials (with the ultraviolet the least
mature regime), and examining arrayed-waveguide-grating spectrographs in
detail, including the phase-error, cross-dispersion, throughput, and
polarization challenges that must be solved to reach the resolving powers HWO
science requires.

Two conclusions stand out. First, the optical and near-infrared photonic
technologies are comparatively mature (TRL 4--5); their path to HWO is one of
optimisation, integration, and qualification. Ultraviolet PICs, by contrast,
require dedicated materials development and are the least mature regime
(TRL 2--3), yet ultraviolet access is the single most-requested capability in
the science-case library. Second, space qualification - radiation
tolerance, thermal cycling, vacuum compatibility, and survival of launch
loads - is the most uniformly under-addressed area and the longest-lead
activity; it should begin immediately, in parallel with optical-performance
development, rather than after it. A phased, fifteen-year programme of
materials and component development, subsystem integration and environmental
qualification, and a technology-demonstrator flight can deliver the required
photonic subsystems at TRL 6 in time for HWO's instrument-definition gate.

\section*{Competing Interests}
The author declares no conflicts of interest.

\noindent\textbf{AI tool disclosure.}
Table formatting, language, grammar, and citation checks were done with the assistance of Grammarly and Claude (Anthropic). All technical content, scientific judgments,
figures, and data presented in the paper are the sole responsibility of the author.

\section*{Data Availability}
No new observational data were generated in the course of this work. The Science Case
Development Documents on which Section~\ref{sect:drivers} draws are openly
accessible through the Astronomical Society of the Pacific~\citep{ASP542} and
as arXiv preprints, and the aggregate science-case statistics are those of
\citet{Dressing2026SCDD}. All other material discussed is available in the
cited literature.

\section*{Acknowledgements}
The author gratefully acknowledges the HWO Science, Technology, Architecture
Review Team (START), the four HWO Community Science Working Groups, and the
members of the Astrophotonics (innoFSPEC) Potsdam group.
This work draws on the open-access HWO25 Proceedings published by the
Astronomical Society of the Pacific. The author thanks the broader HWO
community for the substantial effort invested in preparing these valuable
reference documents.

\noindent\textbf{Funding.} Partial support was provided by the PICS4SENS project, funded by the State of Brandenburg through the Investitionsbank des Landes Brandenburg (ILB), with
support from the European Regional Development Fund (ERDF/EFRE), grant
number 86000879.


\bibliographystyle{rasti}
\bibliography{report}

\bsp	
\label{lastpage}
\end{document}